\documentclass[journal=jpclcd,manuscript=article]{achemso}

\usepackage[version=3]{mhchem} 
\usepackage{braket}
\usepackage{amsmath,amssymb}
\usepackage{physics}
\usepackage{amsmath}
\usepackage{bbold}
\usepackage{braket}
\usepackage{xcolor}
\usepackage{graphicx}
\usepackage{caption}
\usepackage{subcaption}
\usepackage{placeins}
\usepackage{hyperref}

\author{Shreyas Malpathak}
\affiliation[Cornell University]{Department of Chemistry and Chemical Biology, Baker Laboratory, Cornell University Ithaca,14853 NY, USA}
\author{Nandini Ananth}
\affiliation[Cornell University]{Department of Chemistry and Chemical Biology, Baker Laboratory, Cornell University Ithaca,14853 NY, USA}
\email{ananth@cornell.edu}

\title[NO]
  {A Linearized Semiclassical dynamics study of the multi-quantum vibrational relaxation of NO scattering from a Au(111) Surface}

\abbreviations{}

\begin{document}

\begin{tocentry}
\includegraphics[width = 50mm]{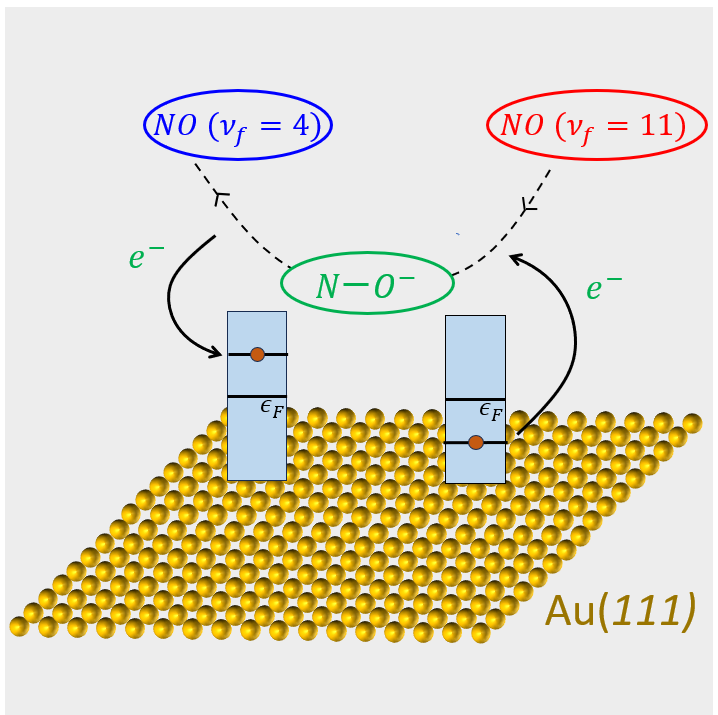}
\end{tocentry}

\begin{abstract}
 The vibrational relaxation of NO molecules scattering from an Au(111) surface has served as the focus of efforts to understand nonadiabatic energy transfer at metal-molecule interfaces. Experimental measurements and previous theoretical efforts suggest that multi-quantal NO vibrational energy relaxation occurs via electron hole pair excitations in the metal. Here, using a Linearized Semiclassical approach, we accurately predict the vibrational relaxation of NO from $\nu_i=3$ state for different incident translational energies. We also accurately capture the central role of transient electron transfer from the metal to the molecule in mediating vibrational relaxation process, but fall short of quantitatively predicting the full extent of multi-quantum relaxation for high incident vibrational excitations ($\nu_i = 16$). 
 
\end{abstract}

The nonadiabatic energy transfer from molecules to metal surfaces has been a topic of much interest in the last couple of decades, with potential applications to heterogeneous catalysis.
~\cite{Tully2000,Wodtke2004,Wodtke2008,Rahinov2011,Golibrzuch2015,Wodtke2016,Rittmeyer2018,Park2019,Jiang2019,Dou2020,Auerbach2021,Zhou2022} The breakdown of the 
Born-Oppenheimer approximation allows nuclear degrees of freedom, especially molecular vibrations, to couple to the continuum of electronic states of the metal. 
Metal-molecule scattering experiments have significantly increased our understanding of these processes
while also raising several interesting questions regarding the mechanism.~\cite{Golibrzuch2015,Wodtke2016,Rittmeyer2018,Park2019} 
 In particular, experiments quantifying the vibrational relaxation of NO molecules scattering from Au(111),\cite{Huang2000,Huang2000b,Li2002,Katz2005,White2005,White2006,Shenvi2006,Shenvi2009,Shenvi2009b,Roy2009,Monturet2010,LaRue2011,Bartels2013,Golibrzuch2013,Golibrzuch2014,Bartels2014,Bartels2014b,Kruger2015,Miao2017,Yin2019,Serwatka2020,Yin2021,Zhang2022,Box2021,Gardner2023b,Gardner2023} and more recently Ag(111),\cite{Kruger2016,Steinsiek2018,Gardner2023} have provided fertile ground 
for theoretical study.
Vibrationally hot NO molecules have been shown to lose multiple quanta of vibrational energy upon inelastic scattering from an Au(111) surface.\cite{Huang2000,Huang2000b,Bartels2013,Golibrzuch2013,Golibrzuch2014,Bartels2014,Bartels2014b,Kruger2015} When the gold surface is doped to lower its work function, the loss of vibrational energy from the NO molecules can even promote the ejection of an electron from the metal surface.\cite{White2005,White2006} These experiments demonstrate the strong coupling between NO vibrations and electron-hole pair (EHP) excitations in the metal, however, to-date theoretical efforts to capture these
effects have had limited success.

To simulate these experiments, approximate quantum dynamical methods, that can
describe nonadiabatic effects are necessary. Early efforts to predict multi-quantum vibrational relaxation relied on  Independent Electron Surface Hopping (IESH)\cite{Shenvi2009,Shenvi2012} and Molecular Dynamics with Electronic Friction (MDEF) methods.\cite{Head-Gordon1995} IESH is a variant of the standard surface hopping approach modified to include a continuum of metal states. MDEF accounts for the effect of the metal states through a friction term in the dynamics, and usually works well when nonadiabatic effects are small. Initial studies with these approaches failed to predict both the dependence of vibrational relaxation on incident translational energy for low incident vibrational excitation,\cite{Golibrzuch2014} and the significant multi-quantum vibrational relaxation seen in experiments for higher incident vibrational excitation.\cite{Kruger2015} These discrepancies between theory and experiment were attributed to deficiencies in the potential energy surface used in these calculations.\cite{Kruger2015} 

More recently, constrained DFT calculations have been used to build a $2 \times 2$ diabatic potential to describe the NO-Au(111) system.\cite{Meng2022} Novel quantum dynamical approaches have also been developed 
to include non-Markovian effects and account for the friction tensor in MDEF methods.\cite{Box2021} A new approach, valid for both weak and strong metal-molecule coupling limits, the Broadened Classical Master Equation (BCME) has also been developed.\cite{Dou2016,Dou2017} In spite of these advances, simulations have not been able to predict all aspects of experimental results. All of these methods 
underestimate multi-quantum relaxation from NO molecules initially in a low energy vibrational state, $\nu_i=3$, to a final vibrational state, $\nu_f = 1$.\cite{Box2021,Gardner2023} Further, 
MDEF and IESH underestimate multi-quantum relaxation for higher incident vibrational states $\nu_i = 11 $ and $16$ as well.\cite{Box2021,Gardner2023} The BCME approach has had some success in reproducing experiments for the high incident vibrational state, $\nu_i = 16$.\cite{Gardner2023}

In this letter, we employ semiclassical (SC) dynamics that have been 
shown to capture quantum effects in dynamics with classical trajectories. 
SC methods are, broadly, based on a stationary phase approximation to the exact path integral formulation of the real-time propagator, and are capable of describing almost all quantum effects \textemdash\, zero-point energy, shallow tunneling, interference and nonadiabatic effects.\cite{Miller2001a,Malpathak2022} In the SC framework, a hierarchy of methods exist that differ by the extent to which quantum effects are captured. Recently, the Mixed Quantum Classical (MQC)-SC method was introduced to filter phase contributions from different degrees of freedom to different extents, allowing for sensitive control of how much each degree of freedom in a complex system is quantized.~\cite{Antipov2015,Church2017,Church2018,Church2019a,Malpathak2022,Malpathak2023} 
The most computationally efficient SC methods are classical-limit approaches like the Linearized Semiclassical (LSC) method, which lack phase information (and consequently cannot capture interference effects) but can incorporate zero-point effects and shallow tunneling.
~\cite{Malpathak2023,Wang1998,Sun1998, Miller2001a} 
LSC-based simulations have been successfully used in describing condensed phase processes,~\cite{Liu2015,Poulsen2005b} and working with the mapping Hamiltonian,~\cite{Meyer1979,Stock1997} LSC methods have also been used to simulate nonadiabatic dynamics with a surprising degree of accuracy.~\cite{Sun1998b,Wang1999,Rabani1999,Ananth2007,Miller2009, Church2018, Miyazaki2023} Here, we investigate the ability of nonadiabatic LSC simulations 
to capture the multi-quantum vibrational relaxation of NO upon inelastic 
scattering from Au(111). We show that this classical-limit SC method is able to capture the 
central role of metal EHPs in the vibrational relaxation of NO and correctly describe cases 
where the incident NO molecule is initially in a low energy vibrational state. However, we find
that, for high incident vibrational states, this method cannot quantitatively reproduce 
the extent of de-excitation observed experimentally.

In modeling the NO-metal potential energy surface, 
we follow previous efforts using a simplified Newns-Anderson-Holstein (NAH) model\cite{Newns1969,Anderson1961,Holstein1959} that represent the metal continuum 
as a discrete set of states that couple to the electronic states of an 
NO molecule incident on the Au(111) surface,
\begin{align}
    \hat{H}_{NAH}(\hat{\mathbf{X}},\hat{\mathbf{P}}) & = \frac{1}{2}\hat{\mathbf{P}}.\mathbf{M}^{-1}.\hat{\mathbf{P}} + U_{0}(\hat{\mathbf{X}}) + h(\hat{\mathbf{X}})\hat{d}^{\dag}\hat{d} + \Sigma_{k=1}^{N} \epsilon_{k}\hat{c}_{k}^{\dag}\hat{c}_{k} \notag  \\ 
    & + \Sigma_{k=1}^{N} V_{k}(\hat{\mathbf{X}})(\hat{d}^{\dag}\hat{c}_{k}+\hat{c}_{k}^{\dag}\hat{d}). \label{eq:NAH-ham}
\end{align}
In Eq.~\eqref{eq:NAH-ham}, $\hat{\mathbf{X}}$, $\hat{\mathbf{P}}$  are vectors corresponding to nuclear position and momenta respectively and $\mathbf{M}$ is a diagonal matrix of the mass of the corresponding dof, the index $k$ runs over a total of $N$ metal states each with energy $\epsilon_k$, and $\hat{d}^{\dag} (\hat{d})$ and $\hat{c}_{k}^{\dag} (\hat{c}_{k})$ are the creation (annihilation) operators for the molecular state and the $k^{th}$ metal state respectively. Finally, in Eq.~\eqref{eq:NAH-ham}, $U_{0}(\hat{\mathbf{X}})$ and $U_{1}(\hat{\mathbf{X}})$ are the potentials corresponding to the neutral NO and NO$^{-}$ state of the molecule. When an electron transfers from the metal states to the NO molecule, $h(\hat{\mathbf{X}}) = U_1(\hat{\mathbf{X}})-U_{0}(\hat{\mathbf{X}})$ serves to change the molecular potential to the NO$^{-}$ state. 
The metal-molecule coupling $V_{k}(\hat{\mathbf{X}})$ is obtained by 
discretizing the hybridization function,~\cite{Shenvi2008,deVega2015,Gardner2023b} 
\begin{align}
\Gamma(\hat{\mathbf{X}},\epsilon)~=~2\pi\Sigma_{k=1}^{N} |V_{k}(\hat{\mathbf{X}})|^{2}\delta(\epsilon - \epsilon_{k}).
\label{eq:gamma}
\end{align}
Following the recently introduced Gardner-Habershon-Maurer (GHM) model,~\cite{Gardner2023} we work with just two nuclear coordinates to capture essential features of the problem -- the NO bond length, $R$, and the distance between the NO center of mass and the surface, $Z$. 
The diabatic potentials $U_{0}(\mathbf{X})$ and  $U_{1}(\mathbf{X})$ and the coupling elements $V_{k}(\mathbf{X})$ are assumed to have functional
form
\begin{align}
    U_0(R,Z) &= V_{M}(R-R_0;D_0,a_0) + \text{exp}(-b_0(Z - Z_0)) + c_0, \label{U0}\\
    U_1(R,Z) &= V_{M}(R-R_1;D_1,a_1) + V_{m}(Z-Z_1;D_2,a_2) + c_1, \label{U1}\\
    V_{k}(Z) &= \bar{V}_k(1-\text{tanh}(Z/\Tilde{a})), \label{Vk}
\end{align}
where $V_{M}(R;D,a) = D[\text{exp}(-2ax) - 2\text{exp}(-ax)]$ is a Morse potential with parameters
$D$ and $a$ determined by fitting to data obtained from constrained DFT calculations.~\cite{Meng2022}
In Eq.~\eqref{Vk}, the strength of the coupling is determined by $\bar{V}_k = \sqrt{\frac{\Gamma}{2\pi}}w_{k}$ where $\Gamma$ is defined in Eq.~\eqref{eq:gamma}, $w_k = \sqrt{\Delta E/N}$,
and $\Delta E$ is the band width of the metal.\cite{Gardner2023b,Shenvi2008} 
In the GHM model, the coupling strength $\Gamma$ is obtained by fitting the ground state adiabatic energy of the NAH Hamiltonian to the ground state energy obtained from reference DFT calculations, while using a wide band of $\Delta E$ = 100 eV. In this work, we refit the model, following the same procedure as in Ref.\citenum{Gardner2023} to obtain the ground state energy of the NAH Hamiltonian, but with a more realistic band width for gold, $\Delta E$ = 7 eV.\cite{Ramchandani1970,Shenvi2009} 
Upon refitting, we obtain a coupling strength of $\Gamma = 3.5 $ eV. More details about the refitting procedure can be found in the supplementary information, and numerical values for all parameters in Eqs.~\eqref{U0}-\eqref{Vk} can be found in Table 1 of Ref.\citenum{Gardner2023}.

The nonadiabatic LSC approach employs a classical analog Hamiltonian obtained by 
using the Meyer-Miller-Stock-Thoss (MMST) mapping to replace both
the bosonic and fermionic creation and annihilation operators by continuous Cartesian operators.\cite{Meyer1979,Stock1997} We note that while the MMST 
mapping has been used successfully in SC simulations of nonadiabatic dynamics involving bosonic
operators,~\cite{Kim2012,Lee2016,Polley2021,Polley2022,Miyazaki2023} recent work has extended the applicability of this mapping 
to systems with non-interacting fermionic states.~\cite{Sun2021,Montoya2023,Jung2023}
The classical NAH Hamiltonian in the MMST framework (symmetrized form) is then,
\begin{align}
    {H}_{sym}({\mathbf{X}},{\mathbf{P}},{\mathbf{x}},{\mathbf{p}}) &= \frac{1}{2}{\mathbf{P}}.\mathbf{M}^{-1}.{\mathbf{P}} + \tilde{U}({\mathbf{X}}) + \frac{1}{2}\left[\mathbf{x}.\mathbf{\tilde{V}}({\mathbf{X}}).\mathbf{x} + \mathbf{p}.\mathbf{\tilde{V}}({\mathbf{X}}).\mathbf{p} \right], \label{h-sym}
\end{align}
where $\tilde{U}({\mathbf{X}}) = U_0({\mathbf{X}}) + \frac{N_e}{N+1}\text{Tr}[V({\mathbf{X}})]$ is the state independent potential and $\mathbf{\tilde{V}}({\mathbf{X}}) =\mathbf{V}({\mathbf{X}}) - \frac{1}{N+1}\text{Tr}[V({\mathbf{X}})]\mathbb{1}$ is the trace-less potential energy matrix. $N_e$ is the total occupation number for the metal states, and $\mathbf{x},\mathbf{p}$ are vectors associated with the position and momentum variables for molecular state metal states. Further details about the symmetrized Hamiltonian, along with a definition of the potential energy matrix $\mathbf{V}(\mathbf{X})$ can be found in the supplementary information.


Within the LSC formalism, for a system with $F$ nuclear dofs and $ D = N+1$ electronic dofs, the expectation value of an operator is, 
\begin{align}
     \langle \hat{B}(t)\rangle  = \frac{1}{(2\pi\hbar)^{F+D}} \int d\mathbf{X}_0\int d\mathbf{P}_0\int d\mathbf{x}_0\int d\mathbf{p}_0 \,\rho_W(\mathbf{X}_0,\mathbf{P}_0,\mathbf{x}_0,\mathbf{p}_0) B_W(\mathbf{X}_t,\mathbf{P}_t,\mathbf{x}_t,\mathbf{p}_t),
     \label{bt}
\end{align}
where $\hat{\rho}$ is the initial density operator, $[.]_W$ indicates the Wigner transform of an operator and the phase space variables at time $t$ are obtained by time-evolving initial phase space variables 
under the symmetrized classical Hamiltonian $H_{sym}$.
The initial density is,
$\hat{\rho} = \hat{\rho}_{\text{Nuc}}\otimes \hat{\rho}_{\text{mol}} \otimes \hat{\rho}_{\text{metal}}$,
where $\hat{\rho}_{\text{Nuc}} \equiv \ketbra{P_{Zi},Z_i}{P_{Zi},Z_i}\otimes\ketbra{\nu_i}{\nu_i} $ is the density operator for the nuclear dofs with the  translational mode, $Z$ being projected onto a coherent state centered at $ (P_{Zi},Z_i) $ and the vibrational dof, $R$ projected on the $\nu_i^{th}$ vibrational state of the NO molecule. $\hat{\rho}_{\text{mol}}$ corresponds to the initially unoccupied anionic molecular state, whereas $\hat{\rho}^{eq}_{\text{metal}}$ corresponds to the occupation of metal states at chemical and thermal equilibrium. Details about the sampling procedures from the initial density are provided in the supplementary information. The experimental observable here is the probability of the NO molecule being in a specific vibrational state $\nu_f$ after
inelastic scattering from the surface. As such, operator $\hat B$ in Eq.~\ref{bt} is $\ketbra{\nu_f}{\nu_f}$. Other operators used in this study, along with their Wigner transforms are presented in the supplementary information. 


In Fig.~\ref{fig:mechNu3}(a), we show the distance of closest approach of the NO $(\nu_i=3)$ molecule to the metal surface decreases steadily with increasing incident energy, $E_i$. This is accompanied by a significant increase in the the extent of electron transfer, as evidenced by the increase in population of the $NO^-$ state, in Fig.~\ref{fig:mechNu3}(b). Examining the underlying 2-D diabatic potential surfaces, $U_0(R,Z)$ and $U_1(R,Z)$, 
we find that when NO is initially in $\nu_i=3$ state, the accessible $R$ values are relatively small and 
as a result, the NO$^-$ state is only energetically accessible when the molecule is very 
close to the metal surface (within $\approx 1.5$\AA). Increasing the incident
translational energy allows the NO molecule to achieve shorter $Z$ distances, enabling significant
charge transfer.
\begin{figure}
    \centering
    \includegraphics[width=0.95\textwidth]{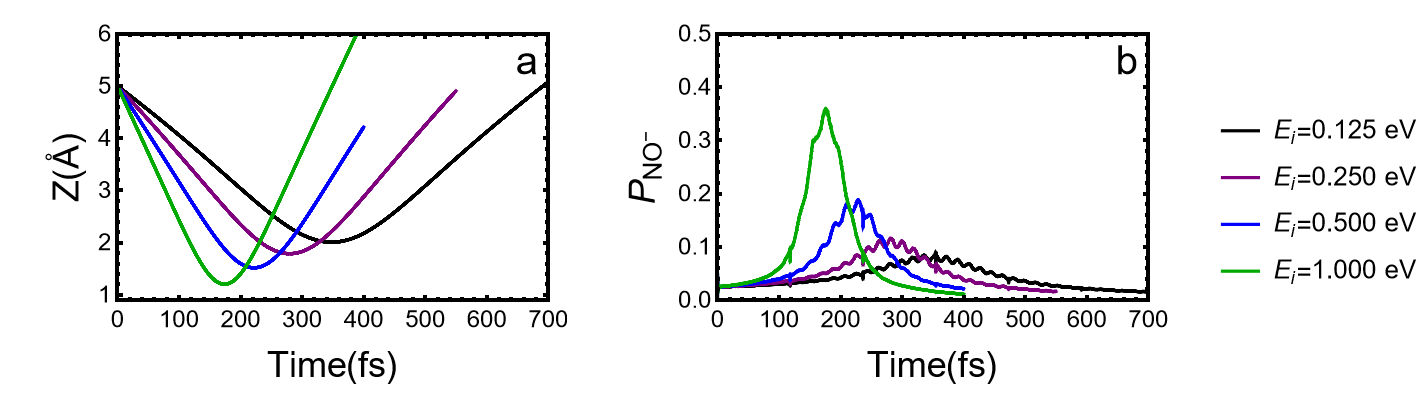}
    \caption{(a) The average distance of NO center of mass from the metal surface (Z) from LSC simulations and (b) the population of the NO$^-$ state ($P_{NO^{-}}$), are plotted as a function of time for $\nu_i =3 $. We show the results for different 
    incident translational energies $E_i = 0.125$ eV (black), $E_i = 0.25$ eV (purple), $E_i = 0.5$ eV (blue), and $E_i = 1.0$ eV (green). } 
    \label{fig:mechNu3}
\end{figure}

\begin{figure}
    \centering
    \includegraphics[width=0.95\textwidth]{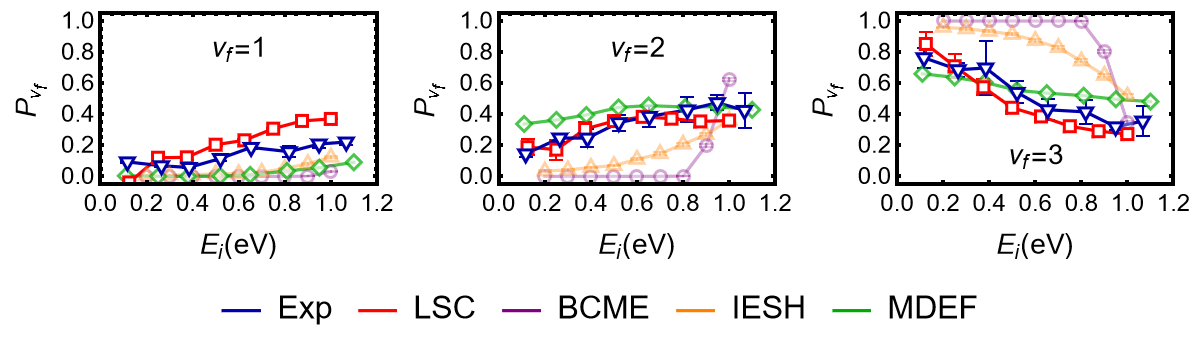}
    \caption{The probability of finding NO in different final vibrational states, $(P_{\nu_f})$, with $\nu_f = 1$, $2$, and $3$ are shown as a function of incident translational energy, $E_i$ for NO initially in $\nu_i=3$ state. We compare results obtained using LSC (red) with previous theoretical results, BCME\cite{Gardner2023} (purple), IESH\cite{Gardner2023} (orange), and MDEF\cite{Box2021} (green), as well as with experimental results\cite{Golibrzuch2014} (blue). Data used to plot the MDEF, BCME, and IESH were taken from Refs. \citenum{Box2021} and \citenum{Gardner2023} as published, and the data showing experimental results was obtained via private communication from the authors of \citenum{Golibrzuch2014}.} 
    \label{fig:Nui3-edep}
\end{figure}

Fig.~\eqref{fig:Nui3-edep} compares the probability of the NO molecule being in final vibrational states $1 \le \nu_f \le 3$ as a function of the incident translational energy for $\nu_i = 3$.
In excellent agreement with experiments,\cite{Golibrzuch2014,Bartels2014} we find 
that the survival probability for $\nu_f =3$ state steadily decreases with increasing incident energy, while the probabilities of transition to $\nu_f = 2$ and to $\nu_f=1$ increase steadily. We also compare
our results with other approximate quantum dynamical approaches in Fig.~\eqref{fig:Nui3-edep} and find that the LSC results are most consistent with experiment, with the other methods like MDEF,~\cite{Box2021}, IESH, and BCME,~\cite{Gardner2023} consistently underestimating multi-quantum relaxation to $\nu_f=1$. 
Together, the results in Fig.~\ref{fig:mechNu3} and Fig.~\ref{fig:Nui3-edep} show that incident translational energy helps overcome the barrier to electron transfer that exists due 
to the inability of NO in low initial vibrational states to access stretched configurations.~\cite{Huang2000,Bartels2014}

\begin{figure}
    \centering
    \includegraphics[width=0.95\textwidth]{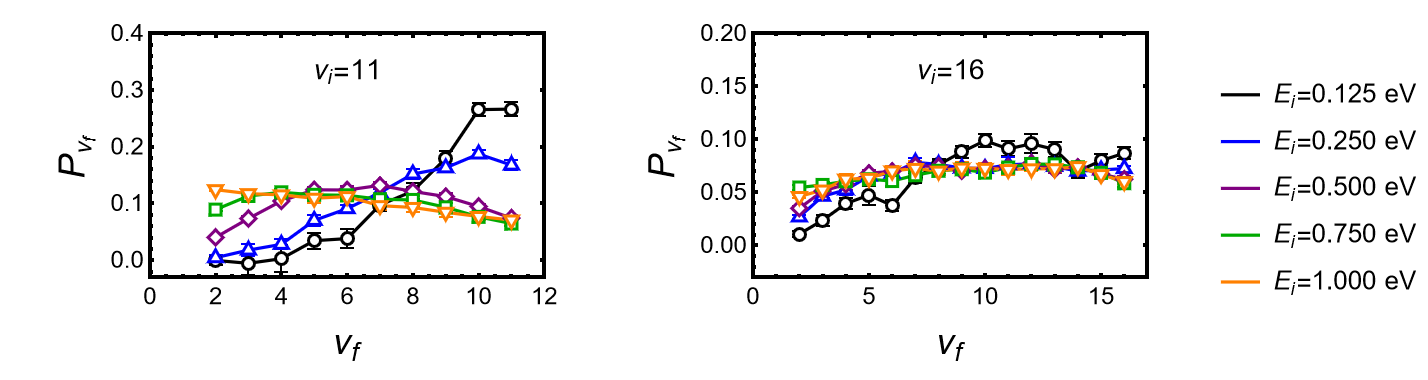}
    \caption{The LSC probability of finding NO in different final vibrational states, $P_{\nu_f}$, is shown
    as a function of final vibrational state $\nu_f$ for $\nu_i = 11$ (left) and $\nu_i=16$ (right). The different colors represent different incident energies, $(E_i)$.} 
    \label{fig:Nui-high-edep}
\end{figure}

Moving forward, we examine the results for NO in higher incident vibrational states, $\nu_i = 11$ and $\nu_i=16$. We find that for both initial conditions, the extent of vibrational relaxation is dependent on incident energy for low incident energies ($E_i = 0.125-0.25$ eV), but becomes more or less independent at higher incident energies as shown in Fig.~\ref{fig:Nui-high-edep}. This does not agree with experimental observations where vibrational relaxation at higher $\nu_i$ was found to be independent of the incident energy.\cite{Bartels2014} Physically, this is reasonable as for $\nu_i \ge 11$ it is expected that large NO bond lengths are energetically accessible, significantly lowering the barrier to electron transfer.  We note that the three other approximate dynamical methods that employ the GHM model for $\nu_i = 16$  all demonstrate similar behavior suggesting that the mismatch with experiment may arise from 
the model potential energy surface.

\begin{figure}
    \centering
    \includegraphics[width=0.95\textwidth]{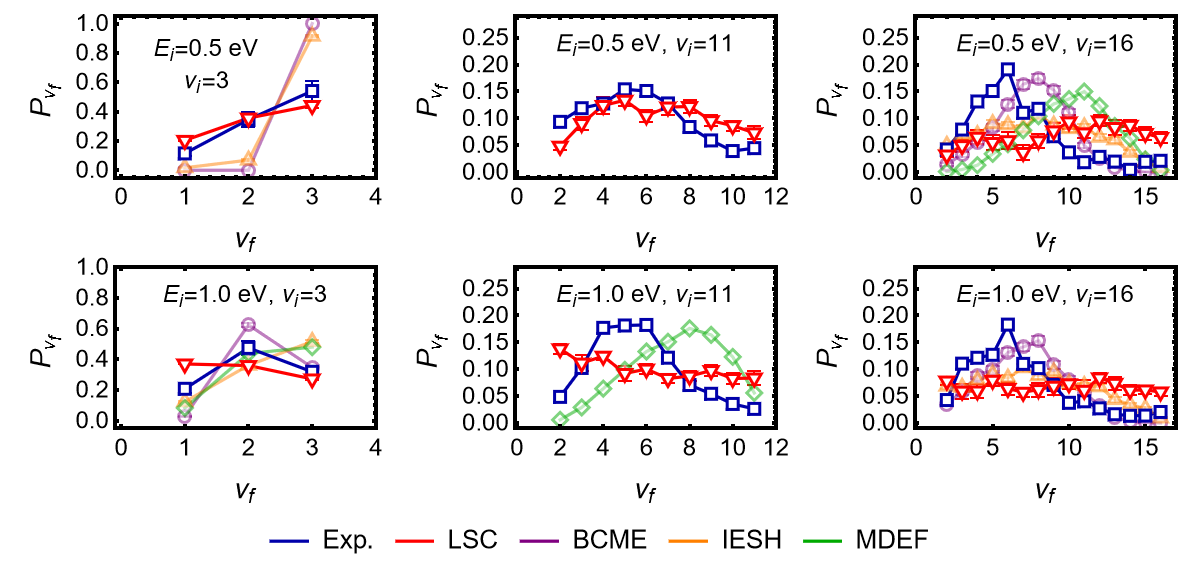}
    \caption{The final vibrational state probability for NO after scattering, $(P_{\nu_f})$, is shown as a function of final vibrational state $\nu_f$ for $\nu_i = 3, 11$ and $16$ for incident energies $E_i = 0.5$ eV (top panel) and $1.0$ eV (bottom panel). We compare the results from the present work using LSC (red), with previous theoretical work BCME~\cite{Gardner2023} (purple), IESH~\cite{Gardner2023} (orange), and MDEF~\cite{Box2021} (green), and with experimental results~\cite{Kruger2015} (blue). Plotted data for MDEF, BCME, and IESH  were taken from Refs.~\citenum{Box2021} and \citenum{Gardner2023} as published, and that for experimental results was obtained via private communication with the authors of Ref.~\citenum{Kruger2015}.}
    \label{fig:nuf}
\end{figure}

Theoretical efforts, thus far, have been unable to consistently and accurately predict the extent of multi-quantum vibrational relaxation and therefore the final vibrational state distribution of NO molecules incident with high vibrational energy $(\nu_i = 11$ and $16$). Initial attempts with electronic friction methods and IESH predicted narrow distributions peaked around the initial vibrational state.\cite{Kruger2015} The discrepancy was attributed to the poor quality of the potential energy surface used, but IESH simulation with the present GHM model also fail to match experimental results.\cite{Kruger2015} More recent MDEF calculations predict some inelastic behavior but fail to capture the full extent of multi-quantum relaxation.\cite{Box2021}  In Fig.~\ref{fig:nuf} we show the distribution of final vibrational states for all three initial vibrational states $\nu_i = 3, 11$ and 16, for two incident energies $E_i = 0.5$ eV and $1.0$ eV. For $\nu_i = 3$, for both incident energies considered, we see that LSC significantly outperforms other theoretical efforts that predict little to no probability of $\nu_f=1$. LSC results for $\nu_i = 11$ are also in very good agreement with experimental results, and show the final vibrational state distribution peaking in the range $\nu_f = 4 -8$ for low incident energy. At higher incident translational energy ($E_i=1$ eV) the agreement is less good although we continue to see multiquantal relaxation being predicted.  

Unfortunately, our results for $\nu_i = 16$ are more sobering, with LSC predicting a very wide, almost uniform probability for all final vibrational states considered here. Interestingly, these results parallel the IESH results with the GHM model.\cite{Gardner2023} And although the wide distributions predict more multi-quantum relaxation than MDEF, both LSC and IESH are outperformed by BCME. It is unclear whether these discrepancies arise due to the approximate nature of the dynamics, the inaccuracies in the fitted model potential, or a shortcoming of the NAH model Hamiltonian in its current form. To explore model-dependence (within the constraints of the current functional form), we verified that both using an improved fit to the NO diabatic states at larger bond lengths and introducing a bond-length ($R$) dependence in the metal-molecule coupling did not change the LSC results. 

\begin{figure}
     \centering
     \begin{subfigure}[b]{0.49\textwidth}
         \centering
         \includegraphics[width=\textwidth]{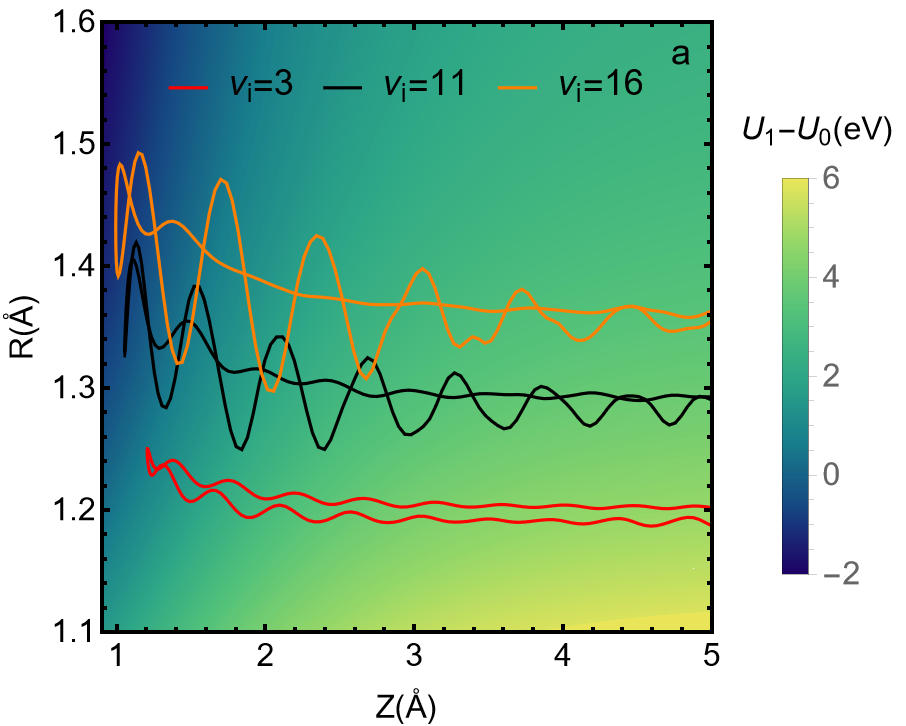}
         \end{subfigure}
     \hfill
     \begin{subfigure}[b]{0.49\textwidth}
         \centering
         \includegraphics[width=\textwidth]{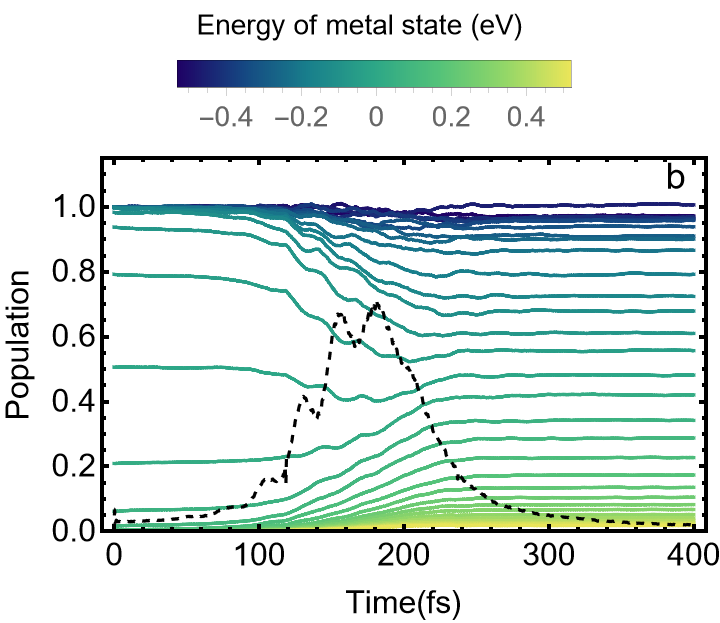}
     \end{subfigure}
     \caption{From the LSC trajectory ensemble for $E_i=1$ eV, we plot the (a) average both length, $R$, and the average distance from the surface, $Z$, as function of time to capture the scattering event overlaid on a contour plot of the potential energy surface, $U_1 - U_0$. Results are shown for three different incident NO vibrational states, $\nu_i = 3$ (red), $\nu_i = 11$ (black) and, $\nu_i = 16$ (orange). The parts of the trajectory with small oscillations in the average bond length correspond to the molecule approaching the metal surface, whereas the parts with larger oscillations correspond to molecule after scattering. (b) For $\nu_i = 16$ and $E_i = 1.0$ eV, we plot the population of a select few metals states (coloured, solid) and the NO$^-$ state (black, dashed) as a function of time. For the metal states, the colour of the line represents the energy of the metal state $(\epsilon_k)$ relative to the Fermi level ($\mu = 0$ eV), as shown in the legend.} \label{fig:pops}
\end{figure}

In order to gain a better understanding of the limitations of LSC, we examine the predicted mechanism of the vibrational energy loss process for different $\nu_i$ in our simulations. Initially, the NO molecule is in the $\nu_i^{th}$ vibrational eigenstate. As the molecule approaches the metal surface the underlying potential is modified by interactions with the metal electrons and the molecule is no longer in an eigenstate. This leads to a small amount of population transfer into neighbouring vibrational states of the isolated NO molecule. The change is evidenced by small oscillations in the average bond length of NO molecules incident on the surface as shown in Fig.~\ref{fig:pops}(a). As the NO molecule approaches closer to the surface, it encounters regions where the NO$^-$ diabatic state is similar $(U_1 - U_0 \approx 0)$ or lower $(U_1 - U_0 < 0)$ in energy than the neutral NO diabatic state. In this region, population transfer from metal states with similar energy ($\epsilon_k \approx U_1 - U_0$) is facilitated, and depending on the extent of electron transfer, the molecular potential is dominated by the NO$^-$ interactions. Upon scattering from the surface, the electron in the NO$^-$ state transfers back to energetically accessible metal states with energy $\epsilon_k \approx U_1 - U_0 > 0 $. The different regions of the diabatic surfaces explored by the NO molecule are shown in Fig.~\ref{fig:pops}(a) where we plot the average bond length (R) and the average distance from the metal surface (Z) for three different initial vibrational states $\nu_i$ superimposed on a contour plot of $U_1 - U_0$.

There are two possible mechanisms for electron transfer between the metal and molecule. If the initial population transfer is from a handful of closely spaced metal states below the Fermi level into the NO$^-$ state, followed by a back transfer into similarly closely spaced metal states above the Fermi level, then multiple quanta of vibrational energy are being transferred to a few EHP excitations, suggesting a `resonant' mechanism. On the other hand, if the observed population transfer involves a large 
number of metal states then vibrational relaxation occurs via a `dissipative' mechanism with a large number of low energy EHP excitations. Experimental measurements of the kinetic energy of electrons emitted from a low work function metal coating on the Au(111) surface upon inelastic NO scattering suggest the mechanism at high initial vibrational energies is resonant rather than dissipative.~\cite{White2005,White2006} 
In Fig.~\ref{fig:pops}(b), we plot the ensemble average population as a function of time for the subset of metal states that show significant population change during the scattering process in our LSC simulations. We find that the relevant metal states have energies $\pm 0.5$ eV around the Fermi level, which is at $\mu = 0$ eV. The line color signifies the energy of the metal state, with dark blue representing states below the Fermi level with energy $\epsilon_k\approx - 0.5$ eV , and yellow representing states above the Fermi level with energy $\epsilon_k \approx 0.5$ eV. States with an initial population $> 0.5$ lie below the Fermi level, states with initial population $< 0.5$ lie above the Fermi level, and the Fermi level, as expected, has initial population of 0.5. We also plot the the population of the NO$^-$ state, and show that as the molecule gets closer to the surface starting at $\approx 100$ fs, metal states below the Fermi level begin to transfer population into the NO$^-$ state. After scattering, the NO molecule moves away from the surface, and at $~200$fs we see back electron transfer from the NO$^-$ state to metal states above the Fermi level. It is clear that our simulations do not support a resonant mechanism, rather we see clearly that multiple quanta of vibrational energy are transferred through several low-energy EHP excitations in a dissipative mechanism. 
We further confirm the role of the dissipative mechanism in our simulations 
by examining the case of $\nu_i = 16$, where the experimentally most likely final 
vibrational state of the NO molecule is $\nu_f = 6$, corresponding to 
an energy loss of 1.9 eV. In our LSC simulations, 
electron transfer to the metal states is confined to a band of width 1 eV centered around the Fermi level. As such, no single transition of 1.9 eV is possible, rather LSC simulations record energy transfer through a series of $<1$eV energy losses to multiple metal states. Similarly, for $\nu_i = 11$, the most likely loss channels in experiments correspond to a loss of $\approx 1$ eV. In our LSC simulations this energy loss is only possible from the edges of the active band of metal states that show almost negligible population change, ruling out a resonant mechanism in this case as well. Our analysis, therefore, clearly shows that vibrational energy loss in our LSC simulations occurs via a dissipative mechanism, and the lack of agreement with experiment for higher incident energies is likely due to the inability to describe multiquantal vibrational energy transfer through a resonant mechanim. 

The vibrational relaxation of NO scattering off Au(111) is a challenging model for 
nonadiabatic dynamic simulations that seek to capture the coupling 
between the vibrational motion of the NO molecules and EHP excitations in the metal. 
Here, we demonstrate that SC 
methods offer a promising strategy to investigate the mechanisms of metal-molecule 
coupling, with even the classical-limit LSC approach capturing multiquantal vibrational 
relaxation. 
Specifically, we show that using an NAH-based Hamiltonian, LSC can correctly predict 
the central role of electron transfer (and therefore EHPs) in mediating vibrational 
energy loss from the NO molecule. We show that, in excellent agreement with experimental results,~\cite{Bartels2014} 
our simulations capture the extent of vibrational relaxation for NO molecules in a 
low initial vibrational state, $\nu_i=3$ and its reliance on incident translational energy 
to overcome the electron transfer barrier. Moreover, for this case, LSC outperforms other theoretical approaches by predicting significant transfer into the $\nu_f = 1$ channel. 
For NO in the $\nu_i=11$ state initially, LSC simulations predict a 
significant degree of multi-quantum relaxation, achieving good agreement with experiments for 
low incident translational energy. However, for high incident translational energy and the case of
$\nu_i=16$, we find although we see multi-quantal energy loss, our simulations cannot capture the 
resonant mechanism supported by experiments. Moving forward, we propose to further study of the effectiveness of the present NAH model Hamiltonian in describing the inelastic scattering of NO from Au(111) using exact dynamic methods like recently developed Heirarchy Equations of Motion (HEOM) approach for the study of metal-molecule interactions.~\cite{Xu2019, Dan2023} It would also be beneficial to modify the current model potentials to include the effect of NO orientation,~\cite{Zhang2022} and to employ the MQC-SC methods developed by our group to move beyond classical-limit SC dynamics.

\begin{acknowledgement}
The authors thank Alec Wodtke, Kai Golibrzuch, and Tim Schäfer for providing experimental data from Refs.~\citenum{Golibrzuch2014} and \citenum{Kruger2015} used to generate lines in Figs.~\ref{fig:Nui3-edep} and \ref{fig:nuf}. The authors also thank James Gardener for sharing information about the fitting procedure used in the GHM model, and also for sharing reference cDFT data used in the fit (employed in Figs.~S1 and S2). 
The authors also acknowledge helpful discussions with Roger Loring.
\end{acknowledgement}

\begin{suppinfo}
Details about the refitting procedure, the refit potential and further details about the simulation are in provided in the Supplementary Information. Data for LSC simulations used to make Figs.1-5 is available at \url{https://github.com/AnanthGroup/NO-Au-Scattering}

\end{suppinfo}

\bibliography{bibfile}

\providecommand{\latin}[1]{#1}
\makeatletter
\providecommand{\doi}
  {\begingroup\let\do\@makeother\dospecials
  \catcode`\{=1 \catcode`\}=2 \doi@aux}
\providecommand{\doi@aux}[1]{\endgroup\texttt{#1}}
\makeatother
\providecommand*\mcitethebibliography{\thebibliography}
\csname @ifundefined\endcsname{endmcitethebibliography}  {\let\endmcitethebibliography\endthebibliography}{}
\begin{mcitethebibliography}{80}
\providecommand*\natexlab[1]{#1}
\providecommand*\mciteSetBstSublistMode[1]{}
\providecommand*\mciteSetBstMaxWidthForm[2]{}
\providecommand*\mciteBstWouldAddEndPuncttrue
  {\def\EndOfBibitem{\unskip.}}
\providecommand*\mciteBstWouldAddEndPunctfalse
  {\let\EndOfBibitem\relax}
\providecommand*\mciteSetBstMidEndSepPunct[3]{}
\providecommand*\mciteSetBstSublistLabelBeginEnd[3]{}
\providecommand*\EndOfBibitem{}
\mciteSetBstSublistMode{f}
\mciteSetBstMaxWidthForm{subitem}{(\alph{mcitesubitemcount})}
\mciteSetBstSublistLabelBeginEnd
  {\mcitemaxwidthsubitemform\space}
  {\relax}
  {\relax}

\bibitem[Tully(2000)]{Tully2000}
Tully,~J.~C. Chemical Dynamics at Metal Surfaces. \emph{Annual Review of Physical Chemistry} \textbf{2000}, \emph{51}, 153--178\relax
\mciteBstWouldAddEndPuncttrue
\mciteSetBstMidEndSepPunct{\mcitedefaultmidpunct}
{\mcitedefaultendpunct}{\mcitedefaultseppunct}\relax
\EndOfBibitem
\bibitem[Alec M.~Wodtke and Auerbach(2004)Alec M.~Wodtke, and Auerbach]{Wodtke2004}
Alec M.~Wodtke,~J. C.~T.; Auerbach,~D.~J. Electronically non-adiabatic interactions of molecules at metal surfaces: Can we trust the Born–Oppenheimer approximation for surface chemistry? \emph{International Reviews in Physical Chemistry} \textbf{2004}, \emph{23}, 513--539\relax
\mciteBstWouldAddEndPuncttrue
\mciteSetBstMidEndSepPunct{\mcitedefaultmidpunct}
{\mcitedefaultendpunct}{\mcitedefaultseppunct}\relax
\EndOfBibitem
\bibitem[Wodtke \latin{et~al.}(2008)Wodtke, Matsiev, and Auerbach]{Wodtke2008}
Wodtke,~A.; Matsiev,~D.; Auerbach,~D. Energy transfer and chemical dynamics at solid surfaces: The special role of charge transfer. \emph{Progress in Surface Science} \textbf{2008}, \emph{83}, 167--214\relax
\mciteBstWouldAddEndPuncttrue
\mciteSetBstMidEndSepPunct{\mcitedefaultmidpunct}
{\mcitedefaultendpunct}{\mcitedefaultseppunct}\relax
\EndOfBibitem
\bibitem[Rahinov \latin{et~al.}(2011)Rahinov, Cooper, Matsiev, Bartels, Auerbach, and Wodtke]{Rahinov2011}
Rahinov,~I.; Cooper,~R.; Matsiev,~D.; Bartels,~C.; Auerbach,~D.~J.; Wodtke,~A.~M. Quantifying the breakdown of the Born–Oppenheimer approximation in surface chemistry. \emph{Physical Chemistry Chemical Physics} \textbf{2011}, \emph{13}, 12680\relax
\mciteBstWouldAddEndPuncttrue
\mciteSetBstMidEndSepPunct{\mcitedefaultmidpunct}
{\mcitedefaultendpunct}{\mcitedefaultseppunct}\relax
\EndOfBibitem
\bibitem[Golibrzuch \latin{et~al.}(2015)Golibrzuch, Bartels, Auerbach, and Wodtke]{Golibrzuch2015}
Golibrzuch,~K.; Bartels,~N.; Auerbach,~D.~J.; Wodtke,~A.~M. The Dynamics of Molecular Interactions and Chemical Reactions at Metal Surfaces: Testing the Foundations of Theory. \emph{Annual Review of Physical Chemistry} \textbf{2015}, \emph{66}, 399--425\relax
\mciteBstWouldAddEndPuncttrue
\mciteSetBstMidEndSepPunct{\mcitedefaultmidpunct}
{\mcitedefaultendpunct}{\mcitedefaultseppunct}\relax
\EndOfBibitem
\bibitem[Wodtke(2016)]{Wodtke2016}
Wodtke,~A.~M. Electronically non-adiabatic influences in surface chemistry and dynamics. \emph{Chemical Society Reviews} \textbf{2016}, \emph{45}, 3641--3657\relax
\mciteBstWouldAddEndPuncttrue
\mciteSetBstMidEndSepPunct{\mcitedefaultmidpunct}
{\mcitedefaultendpunct}{\mcitedefaultseppunct}\relax
\EndOfBibitem
\bibitem[Rittmeyer \latin{et~al.}(2018)Rittmeyer, Bukas, and Reuter]{Rittmeyer2018}
Rittmeyer,~S.~P.; Bukas,~V.~J.; Reuter,~K. Energy dissipation at metal surfaces. \emph{Advances in Physics: X} \textbf{2018}, \emph{3}, 1381574\relax
\mciteBstWouldAddEndPuncttrue
\mciteSetBstMidEndSepPunct{\mcitedefaultmidpunct}
{\mcitedefaultendpunct}{\mcitedefaultseppunct}\relax
\EndOfBibitem
\bibitem[Park \latin{et~al.}(2019)Park, Krüger, Borodin, Kitsopoulos, and Wodtke]{Park2019}
Park,~G.~B.; Krüger,~B.~C.; Borodin,~D.; Kitsopoulos,~T.~N.; Wodtke,~A.~M. Fundamental mechanisms for molecular energy conversion and chemical reactions at surfaces. \emph{Reports on Progress in Physics} \textbf{2019}, \emph{82}, 096401\relax
\mciteBstWouldAddEndPuncttrue
\mciteSetBstMidEndSepPunct{\mcitedefaultmidpunct}
{\mcitedefaultendpunct}{\mcitedefaultseppunct}\relax
\EndOfBibitem
\bibitem[Jiang and Guo(2019)Jiang, and Guo]{Jiang2019}
Jiang,~B.; Guo,~H. Dynamics in reactions on metal surfaces: A theoretical perspective. \emph{The Journal of Chemical Physics} \textbf{2019}, \emph{150}\relax
\mciteBstWouldAddEndPuncttrue
\mciteSetBstMidEndSepPunct{\mcitedefaultmidpunct}
{\mcitedefaultendpunct}{\mcitedefaultseppunct}\relax
\EndOfBibitem
\bibitem[Dou and Subotnik(2020)Dou, and Subotnik]{Dou2020}
Dou,~W.; Subotnik,~J.~E. Nonadiabatic Molecular Dynamics at Metal Surfaces. \emph{The Journal of Physical Chemistry A} \textbf{2020}, \emph{124}, 757--771\relax
\mciteBstWouldAddEndPuncttrue
\mciteSetBstMidEndSepPunct{\mcitedefaultmidpunct}
{\mcitedefaultendpunct}{\mcitedefaultseppunct}\relax
\EndOfBibitem
\bibitem[Auerbach \latin{et~al.}(2021)Auerbach, Tully, and Wodtke]{Auerbach2021}
Auerbach,~D.~J.; Tully,~J.~C.; Wodtke,~A.~M. Chemical dynamics from the gas‐phase to surfaces. \emph{Natural Sciences} \textbf{2021}, \emph{1}, e10005\relax
\mciteBstWouldAddEndPuncttrue
\mciteSetBstMidEndSepPunct{\mcitedefaultmidpunct}
{\mcitedefaultendpunct}{\mcitedefaultseppunct}\relax
\EndOfBibitem
\bibitem[Zhou \latin{et~al.}(2022)Zhou, Meng, Guo, and Jiang]{Zhou2022}
Zhou,~X.; Meng,~G.; Guo,~H.; Jiang,~B. First-Principles Insights into Adiabatic and Nonadiabatic Vibrational Energy-Transfer Dynamics during Molecular Scattering from Metal Surfaces: The Importance of Surface Reactivity. \emph{The Journal of Physical Chemistry Letters} \textbf{2022}, \emph{13}, 3450--3461\relax
\mciteBstWouldAddEndPuncttrue
\mciteSetBstMidEndSepPunct{\mcitedefaultmidpunct}
{\mcitedefaultendpunct}{\mcitedefaultseppunct}\relax
\EndOfBibitem
\bibitem[Huang \latin{et~al.}(2000)Huang, Rettner, Auerbach, and Wodtke]{Huang2000}
Huang,~Y.; Rettner,~C.~T.; Auerbach,~D.~J.; Wodtke,~A.~M. Vibrational Promotion of Electron Transfer. \emph{Science} \textbf{2000}, \emph{290}, 111--114\relax
\mciteBstWouldAddEndPuncttrue
\mciteSetBstMidEndSepPunct{\mcitedefaultmidpunct}
{\mcitedefaultendpunct}{\mcitedefaultseppunct}\relax
\EndOfBibitem
\bibitem[Huang \latin{et~al.}(2000)Huang, Wodtke, Hou, Rettner, and Auerbach]{Huang2000b}
Huang,~Y.; Wodtke,~A.~M.; Hou,~H.; Rettner,~C.~T.; Auerbach,~D.~J. Observation of Vibrational Excitation and Deexcitation for NO $(\nu = 2 )$ Scattering from Au(111): Evidence for Electron-Hole-Pair Mediated Energy Transfer. \emph{Physical Review Letters} \textbf{2000}, \emph{84}, 2985--2988\relax
\mciteBstWouldAddEndPuncttrue
\mciteSetBstMidEndSepPunct{\mcitedefaultmidpunct}
{\mcitedefaultendpunct}{\mcitedefaultseppunct}\relax
\EndOfBibitem
\bibitem[Li and Guo(2002)Li, and Guo]{Li2002}
Li,~S.; Guo,~H. Monte Carlo wave packet study of negative ion mediated vibrationally inelastic scattering of NO from the metal surface. \emph{The Journal of Chemical Physics} \textbf{2002}, \emph{117}, 4499--4508\relax
\mciteBstWouldAddEndPuncttrue
\mciteSetBstMidEndSepPunct{\mcitedefaultmidpunct}
{\mcitedefaultendpunct}{\mcitedefaultseppunct}\relax
\EndOfBibitem
\bibitem[Katz \latin{et~al.}(2005)Katz, Zeiri, and Kosloff]{Katz2005}
Katz,~G.; Zeiri,~Y.; Kosloff,~R. Role of Vibrationally Excited NO in Promoting Electron Emission When Colliding with a Metal Surface: A Nonadiabatic Dynamic Model. \emph{The Journal of Physical Chemistry B} \textbf{2005}, \emph{109}, 18876--18880\relax
\mciteBstWouldAddEndPuncttrue
\mciteSetBstMidEndSepPunct{\mcitedefaultmidpunct}
{\mcitedefaultendpunct}{\mcitedefaultseppunct}\relax
\EndOfBibitem
\bibitem[White \latin{et~al.}(2005)White, Chen, Matsiev, Auerbach, and Wodtke]{White2005}
White,~J.~D.; Chen,~J.; Matsiev,~D.; Auerbach,~D.~J.; Wodtke,~A.~M. Conversion of large-amplitude vibration to electron excitation at a metal surface. \emph{Nature} \textbf{2005}, \emph{433}, 503--505\relax
\mciteBstWouldAddEndPuncttrue
\mciteSetBstMidEndSepPunct{\mcitedefaultmidpunct}
{\mcitedefaultendpunct}{\mcitedefaultseppunct}\relax
\EndOfBibitem
\bibitem[White \latin{et~al.}(2006)White, Chen, Matsiev, Auerbach, and Wodtke]{White2006}
White,~J.~D.; Chen,~J.; Matsiev,~D.; Auerbach,~D.~J.; Wodtke,~A.~M. Vibrationally promoted electron emission from low work-function metal surfaces. \emph{The Journal of Chemical Physics} \textbf{2006}, \emph{124}, 64702\relax
\mciteBstWouldAddEndPuncttrue
\mciteSetBstMidEndSepPunct{\mcitedefaultmidpunct}
{\mcitedefaultendpunct}{\mcitedefaultseppunct}\relax
\EndOfBibitem
\bibitem[Shenvi \latin{et~al.}(2006)Shenvi, Roy, Parandekar, and Tully]{Shenvi2006}
Shenvi,~N.; Roy,~S.; Parandekar,~P.; Tully,~J. Vibrational relaxation of NO on Au(111) via electron-hole pair generation. \emph{The Journal of Chemical Physics} \textbf{2006}, \emph{125}, 154703\relax
\mciteBstWouldAddEndPuncttrue
\mciteSetBstMidEndSepPunct{\mcitedefaultmidpunct}
{\mcitedefaultendpunct}{\mcitedefaultseppunct}\relax
\EndOfBibitem
\bibitem[Shenvi \latin{et~al.}(2009)Shenvi, Roy, and Tully]{Shenvi2009}
Shenvi,~N.; Roy,~S.; Tully,~J.~C. {Nonadiabatic dynamics at metal surfaces: Independent-electron surface hopping}. \emph{The Journal of Chemical Physics} \textbf{2009}, \emph{130}, 174107\relax
\mciteBstWouldAddEndPuncttrue
\mciteSetBstMidEndSepPunct{\mcitedefaultmidpunct}
{\mcitedefaultendpunct}{\mcitedefaultseppunct}\relax
\EndOfBibitem
\bibitem[Shenvi \latin{et~al.}(2009)Shenvi, Roy, and Tully]{Shenvi2009b}
Shenvi,~N.; Roy,~S.; Tully,~J.~C. Dynamical Steering and Electronic Excitation in NO Scattering from a Gold Surface. \emph{Science} \textbf{2009}, \emph{326}, 829--832\relax
\mciteBstWouldAddEndPuncttrue
\mciteSetBstMidEndSepPunct{\mcitedefaultmidpunct}
{\mcitedefaultendpunct}{\mcitedefaultseppunct}\relax
\EndOfBibitem
\bibitem[Roy \latin{et~al.}(2009)Roy, Shenvi, and Tully]{Roy2009}
Roy,~S.; Shenvi,~N.~A.; Tully,~J.~C. Model Hamiltonian for the interaction of NO with the Au(111) surface. \emph{The Journal of Chemical Physics} \textbf{2009}, \emph{130}, 174716\relax
\mciteBstWouldAddEndPuncttrue
\mciteSetBstMidEndSepPunct{\mcitedefaultmidpunct}
{\mcitedefaultendpunct}{\mcitedefaultseppunct}\relax
\EndOfBibitem
\bibitem[Monturet and Saalfrank(2010)Monturet, and Saalfrank]{Monturet2010}
Monturet,~S.; Saalfrank,~P. Role of electronic friction during the scattering of vibrationally excited nitric oxide molecules from Au(111). \emph{Physical Review B} \textbf{2010}, \emph{82}, 075404\relax
\mciteBstWouldAddEndPuncttrue
\mciteSetBstMidEndSepPunct{\mcitedefaultmidpunct}
{\mcitedefaultendpunct}{\mcitedefaultseppunct}\relax
\EndOfBibitem
\bibitem[LaRue \latin{et~al.}(2011)LaRue, Schäfer, Matsiev, Velarde, Nahler, Auerbach, and Wodtke]{LaRue2011}
LaRue,~J.; Schäfer,~T.; Matsiev,~D.; Velarde,~L.; Nahler,~N.~H.; Auerbach,~D.~J.; Wodtke,~A.~M. Vibrationally promoted electron emission at a metal surface: electron kinetic energy distributions. \emph{Phys. Chem. Chem. Phys.} \textbf{2011}, \emph{13}, 97--99\relax
\mciteBstWouldAddEndPuncttrue
\mciteSetBstMidEndSepPunct{\mcitedefaultmidpunct}
{\mcitedefaultendpunct}{\mcitedefaultseppunct}\relax
\EndOfBibitem
\bibitem[Bartels \latin{et~al.}(2013)Bartels, Golibrzuch, Bartels, Chen, Auerbach, Wodtke, and Schäfer]{Bartels2013}
Bartels,~N.; Golibrzuch,~K.; Bartels,~C.; Chen,~L.; Auerbach,~D.~J.; Wodtke,~A.~M.; Schäfer,~T. Observation of orientation-dependent electron transfer in molecule–surface collisions. \emph{Proceedings of the National Academy of Sciences} \textbf{2013}, \emph{110}, 17738--17743\relax
\mciteBstWouldAddEndPuncttrue
\mciteSetBstMidEndSepPunct{\mcitedefaultmidpunct}
{\mcitedefaultendpunct}{\mcitedefaultseppunct}\relax
\EndOfBibitem
\bibitem[Golibrzuch \latin{et~al.}(2013)Golibrzuch, Shirhatti, Altschäffel, Rahinov, Auerbach, Wodtke, and Bartels]{Golibrzuch2013}
Golibrzuch,~K.; Shirhatti,~P.~R.; Altschäffel,~J.; Rahinov,~I.; Auerbach,~D.~J.; Wodtke,~A.~M.; Bartels,~C. State-to-State Time-of-Flight Measurements of NO Scattering from Au(111): Direct Observation of Translation-to-Vibration Coupling in Electronically Nonadiabatic Energy Transfer. \emph{The Journal of Physical Chemistry A} \textbf{2013}, \emph{117}, 8750--8760\relax
\mciteBstWouldAddEndPuncttrue
\mciteSetBstMidEndSepPunct{\mcitedefaultmidpunct}
{\mcitedefaultendpunct}{\mcitedefaultseppunct}\relax
\EndOfBibitem
\bibitem[Golibrzuch \latin{et~al.}(2014)Golibrzuch, Shirhatti, Rahinov, Kandratsenka, Auerbach, Wodtke, and Bartels]{Golibrzuch2014}
Golibrzuch,~K.; Shirhatti,~P.~R.; Rahinov,~I.; Kandratsenka,~A.; Auerbach,~D.~J.; Wodtke,~A.~M.; Bartels,~C. The importance of accurate adiabatic interaction potentials for the correct description of electronically nonadiabatic vibrational energy transfer: A combined experimental and theoretical study of NO( v= 3) collisions with a Au(111) surface. \emph{The Journal of Chemical Physics} \textbf{2014}, \emph{140}, 044701\relax
\mciteBstWouldAddEndPuncttrue
\mciteSetBstMidEndSepPunct{\mcitedefaultmidpunct}
{\mcitedefaultendpunct}{\mcitedefaultseppunct}\relax
\EndOfBibitem
\bibitem[Bartels \latin{et~al.}(2014)Bartels, Krüger, Auerbach, Wodtke, and Schäfer]{Bartels2014}
Bartels,~N.; Krüger,~B.~C.; Auerbach,~D.~J.; Wodtke,~A.~M.; Schäfer,~T. Controlling an Electron-Transfer Reaction at a Metal Surface by Manipulating Reactant Motion and Orientation. \emph{Angewandte Chemie International Edition} \textbf{2014}, \emph{53}, 13690--13694\relax
\mciteBstWouldAddEndPuncttrue
\mciteSetBstMidEndSepPunct{\mcitedefaultmidpunct}
{\mcitedefaultendpunct}{\mcitedefaultseppunct}\relax
\EndOfBibitem
\bibitem[Bartels \latin{et~al.}(2014)Bartels, Golibrzuch, Bartels, Chen, Auerbach, Wodtke, and Schäfer]{Bartels2014b}
Bartels,~N.; Golibrzuch,~K.; Bartels,~C.; Chen,~L.; Auerbach,~D.~J.; Wodtke,~A.~M.; Schäfer,~T. Dynamical steering in an electron transfer surface reaction: Oriented $NO(\nu = 3, 0.08 < E_i < 0.89 eV)$ relaxation in collisions with a Au(111) surface. \emph{The Journal of Chemical Physics} \textbf{2014}, \emph{140}, 54710\relax
\mciteBstWouldAddEndPuncttrue
\mciteSetBstMidEndSepPunct{\mcitedefaultmidpunct}
{\mcitedefaultendpunct}{\mcitedefaultseppunct}\relax
\EndOfBibitem
\bibitem[Krüger \latin{et~al.}(2015)Krüger, Bartels, Bartels, Kandratsenka, Tully, Wodtke, and Schäfer]{Kruger2015}
Krüger,~B.~C.; Bartels,~N.; Bartels,~C.; Kandratsenka,~A.; Tully,~J.~C.; Wodtke,~A.~M.; Schäfer,~T. NO Vibrational Energy Transfer on a Metal Surface: Still a Challenge to First-Principles Theory. \emph{The Journal of Physical Chemistry C} \textbf{2015}, \emph{119}, 3268--3272\relax
\mciteBstWouldAddEndPuncttrue
\mciteSetBstMidEndSepPunct{\mcitedefaultmidpunct}
{\mcitedefaultendpunct}{\mcitedefaultseppunct}\relax
\EndOfBibitem
\bibitem[Miao \latin{et~al.}(2017)Miao, Dou, and Subotnik]{Miao2017}
Miao,~G.; Dou,~W.; Subotnik,~J. Vibrational relaxation at a metal surface: Electronic friction versus classical master equations. \emph{The Journal of Chemical Physics} \textbf{2017}, \emph{147}, 224105\relax
\mciteBstWouldAddEndPuncttrue
\mciteSetBstMidEndSepPunct{\mcitedefaultmidpunct}
{\mcitedefaultendpunct}{\mcitedefaultseppunct}\relax
\EndOfBibitem
\bibitem[Yin \latin{et~al.}(2019)Yin, Zhang, and Jiang]{Yin2019}
Yin,~R.; Zhang,~Y.; Jiang,~B. Strong Vibrational Relaxation of NO Scattered from Au(111): Importance of the Adiabatic Potential Energy Surface. \emph{The Journal of Physical Chemistry Letters} \textbf{2019}, \emph{10}, 5969--5974\relax
\mciteBstWouldAddEndPuncttrue
\mciteSetBstMidEndSepPunct{\mcitedefaultmidpunct}
{\mcitedefaultendpunct}{\mcitedefaultseppunct}\relax
\EndOfBibitem
\bibitem[Serwatka \latin{et~al.}(2020)Serwatka, Füchsel, and Tremblay]{Serwatka2020}
Serwatka,~T.; Füchsel,~G.; Tremblay,~J.~C. Scattering of $NO(\nu = 3)$ from Au(111): a stochastic dissipative quantum dynamical perspective. \emph{Physical Chemistry Chemical Physics} \textbf{2020}, \emph{22}, 6584--6594\relax
\mciteBstWouldAddEndPuncttrue
\mciteSetBstMidEndSepPunct{\mcitedefaultmidpunct}
{\mcitedefaultendpunct}{\mcitedefaultseppunct}\relax
\EndOfBibitem
\bibitem[Yin and Jiang(2021)Yin, and Jiang]{Yin2021}
Yin,~R.; Jiang,~B. Mechanical Vibrational Relaxation of NO Scattering from Metal and Insulator Surfaces: When and Why They Are Different. \emph{Physical Review Letters} \textbf{2021}, \emph{126}, 156101\relax
\mciteBstWouldAddEndPuncttrue
\mciteSetBstMidEndSepPunct{\mcitedefaultmidpunct}
{\mcitedefaultendpunct}{\mcitedefaultseppunct}\relax
\EndOfBibitem
\bibitem[Zhang \latin{et~al.}(2022)Zhang, Box, Schäfer, Kandratsenka, Wodtke, Maurer, and Jiang]{Zhang2022}
Zhang,~Y.; Box,~C.~L.; Schäfer,~T.; Kandratsenka,~A.; Wodtke,~A.~M.; Maurer,~R.~J.; Jiang,~B. Stereodynamics of adiabatic and non-adiabatic energy transfer in a molecule surface encounter. \emph{Physical Chemistry Chemical Physics} \textbf{2022}, \emph{24}, 19753--19760\relax
\mciteBstWouldAddEndPuncttrue
\mciteSetBstMidEndSepPunct{\mcitedefaultmidpunct}
{\mcitedefaultendpunct}{\mcitedefaultseppunct}\relax
\EndOfBibitem
\bibitem[Box \latin{et~al.}(2021)Box, Zhang, Yin, Jiang, and Maurer]{Box2021}
Box,~C.~L.; Zhang,~Y.; Yin,~R.; Jiang,~B.; Maurer,~R.~J. Determining the Effect of Hot Electron Dissipation on Molecular Scattering Experiments at Metal Surfaces. \emph{JACS Au} \textbf{2021}, \emph{1}, 164--173\relax
\mciteBstWouldAddEndPuncttrue
\mciteSetBstMidEndSepPunct{\mcitedefaultmidpunct}
{\mcitedefaultendpunct}{\mcitedefaultseppunct}\relax
\EndOfBibitem
\bibitem[Gardner \latin{et~al.}(2023)Gardner, Corken, Janke, Al, Habershon, and Maurer]{Gardner2023b}
Gardner,~J.; Corken,~D.; Janke,~S.~M.; Al; Habershon,~S.; Maurer,~R.~J. {Efficient implementation and performance analysis of the independent electron surface hopping method for dynamics at metal surfaces}. \emph{The Journal of Chemical Physics} \textbf{2023}, \emph{158}, 64101\relax
\mciteBstWouldAddEndPuncttrue
\mciteSetBstMidEndSepPunct{\mcitedefaultmidpunct}
{\mcitedefaultendpunct}{\mcitedefaultseppunct}\relax
\EndOfBibitem
\bibitem[Gardner \latin{et~al.}(2023)Gardner, Habershon, and Maurer]{Gardner2023}
Gardner,~J.; Habershon,~S.; Maurer,~R.~J. {Assessing Mixed Quantum-Classical Molecular Dynamics Methods for Nonadiabatic Dynamics of Molecules on Metal Surfaces}. \emph{The Journal of Physical Chemistry C} \textbf{2023}, \relax
\mciteBstWouldAddEndPunctfalse
\mciteSetBstMidEndSepPunct{\mcitedefaultmidpunct}
{}{\mcitedefaultseppunct}\relax
\EndOfBibitem
\bibitem[Krüger \latin{et~al.}(2016)Krüger, Meyer, Kandratsenka, Wodtke, and Schäfer]{Kruger2016}
Krüger,~B.~C.; Meyer,~S.; Kandratsenka,~A.; Wodtke,~A.~M.; Schäfer,~T. Vibrational Inelasticity of Highly Vibrationally Excited NO on Ag(111). \emph{The Journal of Physical Chemistry Letters} \textbf{2016}, \emph{7}, 441--446\relax
\mciteBstWouldAddEndPuncttrue
\mciteSetBstMidEndSepPunct{\mcitedefaultmidpunct}
{\mcitedefaultendpunct}{\mcitedefaultseppunct}\relax
\EndOfBibitem
\bibitem[Steinsiek \latin{et~al.}(2018)Steinsiek, Shirhatti, Geweke, Bartels, and Wodtke]{Steinsiek2018}
Steinsiek,~C.; Shirhatti,~P.~R.; Geweke,~J.; Bartels,~C.; Wodtke,~A.~M. Work Function Dependence of Vibrational Relaxation Probabilities: NO( v = 2) Scattering from Ultrathin Metallic Films of Ag/Au(111). \emph{The Journal of Physical Chemistry C} \textbf{2018}, \emph{122}, 10027--10033\relax
\mciteBstWouldAddEndPuncttrue
\mciteSetBstMidEndSepPunct{\mcitedefaultmidpunct}
{\mcitedefaultendpunct}{\mcitedefaultseppunct}\relax
\EndOfBibitem
\bibitem[Shenvi and Tully(2012)Shenvi, and Tully]{Shenvi2012}
Shenvi,~N.; Tully,~J.~C. {Nonadiabatic dynamics at metal surfaces: Independent electron surface hopping with phonon and electron thermostats}. \emph{Faraday Discussions} \textbf{2012}, \emph{157}, 325\relax
\mciteBstWouldAddEndPuncttrue
\mciteSetBstMidEndSepPunct{\mcitedefaultmidpunct}
{\mcitedefaultendpunct}{\mcitedefaultseppunct}\relax
\EndOfBibitem
\bibitem[Head-Gordon and Tully(1995)Head-Gordon, and Tully]{Head-Gordon1995}
Head-Gordon,~M.; Tully,~J.~C. {Molecular dynamics with electronic frictions}. \emph{The Journal of Chemical Physics} \textbf{1995}, \emph{103}, 10137--10145\relax
\mciteBstWouldAddEndPuncttrue
\mciteSetBstMidEndSepPunct{\mcitedefaultmidpunct}
{\mcitedefaultendpunct}{\mcitedefaultseppunct}\relax
\EndOfBibitem
\bibitem[Meng and Jiang(2022)Meng, and Jiang]{Meng2022}
Meng,~G.; Jiang,~B. {A pragmatic protocol for determining charge transfer states of molecules at metal surfaces by constrained density functional theory}. \emph{The Journal of Chemical Physics} \textbf{2022}, \emph{157}, 214103\relax
\mciteBstWouldAddEndPuncttrue
\mciteSetBstMidEndSepPunct{\mcitedefaultmidpunct}
{\mcitedefaultendpunct}{\mcitedefaultseppunct}\relax
\EndOfBibitem
\bibitem[Dou and Subotnik(2016)Dou, and Subotnik]{Dou2016}
Dou,~W.; Subotnik,~J.~E. {A broadened classical master equation approach for nonadiabatic dynamics at metal surfaces: Beyond the weak molecule-metal coupling limit}. \emph{The Journal of Chemical Physics} \textbf{2016}, \emph{144}, 24116\relax
\mciteBstWouldAddEndPuncttrue
\mciteSetBstMidEndSepPunct{\mcitedefaultmidpunct}
{\mcitedefaultendpunct}{\mcitedefaultseppunct}\relax
\EndOfBibitem
\bibitem[Dou and Subotnik(2017)Dou, and Subotnik]{Dou2017}
Dou,~W.; Subotnik,~J.~E. {Electronic friction near metal surfaces: A case where molecule-metal couplings depend on nuclear coordinates}. \emph{The Journal of Chemical Physics} \textbf{2017}, \emph{146}, 92304\relax
\mciteBstWouldAddEndPuncttrue
\mciteSetBstMidEndSepPunct{\mcitedefaultmidpunct}
{\mcitedefaultendpunct}{\mcitedefaultseppunct}\relax
\EndOfBibitem
\bibitem[Miller(2001)]{Miller2001a}
Miller,~W.~H. The Semiclassical Initial Value Representation: A Potentially Practical Way for Adding Quantum Effects to Classical Molecular Dynamics Simulations. \emph{The Journal of Physical Chemistry A} \textbf{2001}, \emph{105}, 2942--2955\relax
\mciteBstWouldAddEndPuncttrue
\mciteSetBstMidEndSepPunct{\mcitedefaultmidpunct}
{\mcitedefaultendpunct}{\mcitedefaultseppunct}\relax
\EndOfBibitem
\bibitem[Malpathak \latin{et~al.}(2022)Malpathak, Church, and Ananth]{Malpathak2022}
Malpathak,~S.; Church,~M.~S.; Ananth,~N. A Semiclassical Framework for Mixed Quantum Classical Dynamics. \emph{The Journal of Physical Chemistry A} \textbf{2022}, \emph{126}, 6359--6375\relax
\mciteBstWouldAddEndPuncttrue
\mciteSetBstMidEndSepPunct{\mcitedefaultmidpunct}
{\mcitedefaultendpunct}{\mcitedefaultseppunct}\relax
\EndOfBibitem
\bibitem[Antipov \latin{et~al.}(2015)Antipov, Ye, and Ananth]{Antipov2015}
Antipov,~S.~V.; Ye,~Z.; Ananth,~N. Dynamically consistent method for mixed quantum-classical simulations: A semiclassical approach. \emph{The Journal of Chemical Physics} \textbf{2015}, \emph{142}, 184102\relax
\mciteBstWouldAddEndPuncttrue
\mciteSetBstMidEndSepPunct{\mcitedefaultmidpunct}
{\mcitedefaultendpunct}{\mcitedefaultseppunct}\relax
\EndOfBibitem
\bibitem[Church \latin{et~al.}(2017)Church, Antipov, and Ananth]{Church2017}
Church,~M.~S.; Antipov,~S.~V.; Ananth,~N. Validating and implementing modified Filinov phase filtration in semiclassical dynamics. \emph{The Journal of Chemical Physics} \textbf{2017}, \emph{146}, 234104\relax
\mciteBstWouldAddEndPuncttrue
\mciteSetBstMidEndSepPunct{\mcitedefaultmidpunct}
{\mcitedefaultendpunct}{\mcitedefaultseppunct}\relax
\EndOfBibitem
\bibitem[Church \latin{et~al.}(2018)Church, Hele, Ezra, and Ananth]{Church2018}
Church,~M.~S.; Hele,~T. J.~H.; Ezra,~G.~S.; Ananth,~N. Nonadiabatic semiclassical dynamics in the mixed quantum-classical initial value representation. \emph{The Journal of Chemical Physics} \textbf{2018}, \emph{148}, 102326\relax
\mciteBstWouldAddEndPuncttrue
\mciteSetBstMidEndSepPunct{\mcitedefaultmidpunct}
{\mcitedefaultendpunct}{\mcitedefaultseppunct}\relax
\EndOfBibitem
\bibitem[Church and Ananth(2019)Church, and Ananth]{Church2019a}
Church,~M.~S.; Ananth,~N. Semiclassical dynamics in the mixed quantum-classical limit. \emph{The Journal of Chemical Physics} \textbf{2019}, \emph{151}, 134109\relax
\mciteBstWouldAddEndPuncttrue
\mciteSetBstMidEndSepPunct{\mcitedefaultmidpunct}
{\mcitedefaultendpunct}{\mcitedefaultseppunct}\relax
\EndOfBibitem
\bibitem[Malpathak and Ananth(2023)Malpathak, and Ananth]{Malpathak2023}
Malpathak,~S.; Ananth,~N. Non-linear correlation functions and zero-point energy flow in mixed quantum–classical semiclassical dynamics. \emph{The Journal of Chemical Physics} \textbf{2023}, \emph{158}, 104106\relax
\mciteBstWouldAddEndPuncttrue
\mciteSetBstMidEndSepPunct{\mcitedefaultmidpunct}
{\mcitedefaultendpunct}{\mcitedefaultseppunct}\relax
\EndOfBibitem
\bibitem[Wang \latin{et~al.}(1998)Wang, Sun, and Miller]{Wang1998}
Wang,~H.; Sun,~X.; Miller,~W.~H. Semiclassical approximations for the calculation of thermal rate constants for chemical reactions in complex molecular systems. \emph{Journal of Chemical Physics} \textbf{1998}, \emph{108}, 9726--9736\relax
\mciteBstWouldAddEndPuncttrue
\mciteSetBstMidEndSepPunct{\mcitedefaultmidpunct}
{\mcitedefaultendpunct}{\mcitedefaultseppunct}\relax
\EndOfBibitem
\bibitem[Sun \latin{et~al.}(1998)Sun, Wang, and Miller]{Sun1998}
Sun,~X.; Wang,~H.; Miller,~W.~H. On the semiclassical description of quantum coherence in thermal rate constants. \emph{Journal of Chemical Physics} \textbf{1998}, \emph{109}, 4190\relax
\mciteBstWouldAddEndPuncttrue
\mciteSetBstMidEndSepPunct{\mcitedefaultmidpunct}
{\mcitedefaultendpunct}{\mcitedefaultseppunct}\relax
\EndOfBibitem
\bibitem[Liu(2015)]{Liu2015}
Liu,~J. Recent advances in the linearized semiclassical initial value representation/classical Wigner model for the thermal correlation function. \emph{International Journal of Quantum Chemistry} \textbf{2015}, \emph{115}, 657--670\relax
\mciteBstWouldAddEndPuncttrue
\mciteSetBstMidEndSepPunct{\mcitedefaultmidpunct}
{\mcitedefaultendpunct}{\mcitedefaultseppunct}\relax
\EndOfBibitem
\bibitem[Poulsen \latin{et~al.}(2005)Poulsen, Nyman, and Rossky]{Poulsen2005b}
Poulsen,~J.~A.; Nyman,~G.; Rossky,~P.~J. Static and dynamic quantum effects in molecular liquids: A linearized path integral description of water. \emph{Proceedings of the National Academy of Sciences} \textbf{2005}, \emph{102}, 6709\relax
\mciteBstWouldAddEndPuncttrue
\mciteSetBstMidEndSepPunct{\mcitedefaultmidpunct}
{\mcitedefaultendpunct}{\mcitedefaultseppunct}\relax
\EndOfBibitem
\bibitem[Meyer and Miller(1979)Meyer, and Miller]{Meyer1979}
Meyer,~H.-D.; Miller,~W.~H. A classical analog for electronic degrees of freedom in nonadiabatic collision processes. \emph{The Journal of Chemical Physics} \textbf{1979}, \emph{70}, 3214--3223\relax
\mciteBstWouldAddEndPuncttrue
\mciteSetBstMidEndSepPunct{\mcitedefaultmidpunct}
{\mcitedefaultendpunct}{\mcitedefaultseppunct}\relax
\EndOfBibitem
\bibitem[Stock and Thoss(1997)Stock, and Thoss]{Stock1997}
Stock,~G.; Thoss,~M. Semiclassical Description of Nonadiabatic Quantum Dynamics. \emph{Physical Review Letters} \textbf{1997}, \emph{78}, 578\relax
\mciteBstWouldAddEndPuncttrue
\mciteSetBstMidEndSepPunct{\mcitedefaultmidpunct}
{\mcitedefaultendpunct}{\mcitedefaultseppunct}\relax
\EndOfBibitem
\bibitem[Sun \latin{et~al.}(1998)Sun, Wang, and Miller]{Sun1998b}
Sun,~X.; Wang,~H.; Miller,~W.~H. {Semiclassical theory of electronically nonadiabatic dynamics: Results of a linearized approximation to the initial value representation}. \emph{The Journal of Chemical Physics} \textbf{1998}, \emph{109}, 7064\relax
\mciteBstWouldAddEndPuncttrue
\mciteSetBstMidEndSepPunct{\mcitedefaultmidpunct}
{\mcitedefaultendpunct}{\mcitedefaultseppunct}\relax
\EndOfBibitem
\bibitem[Wang \latin{et~al.}(1999)Wang, Song, Chandler, and Miller]{Wang1999}
Wang,~H.; Song,~X.; Chandler,~D.; Miller,~W.~H. {Semiclassical study of electronically nonadiabatic dynamics in the condensed-phase: Spin-boson problem with Debye spectral density}. \emph{The Journal of Chemical Physics} \textbf{1999}, \emph{110}, 4828--4840\relax
\mciteBstWouldAddEndPuncttrue
\mciteSetBstMidEndSepPunct{\mcitedefaultmidpunct}
{\mcitedefaultendpunct}{\mcitedefaultseppunct}\relax
\EndOfBibitem
\bibitem[Rabani \latin{et~al.}(1999)Rabani, Egorov, and Berne]{Rabani1999}
Rabani,~E.; Egorov,~S.~A.; Berne,~B.~J. {Classical Approximation to Nonradiative Electronic Relaxation in Condensed Phase Systems}. \emph{The Journal of Physical Chemistry A} \textbf{1999}, \emph{103}, 9539--9544\relax
\mciteBstWouldAddEndPuncttrue
\mciteSetBstMidEndSepPunct{\mcitedefaultmidpunct}
{\mcitedefaultendpunct}{\mcitedefaultseppunct}\relax
\EndOfBibitem
\bibitem[Ananth \latin{et~al.}(2007)Ananth, Venkataraman, and Miller]{Ananth2007}
Ananth,~N.; Venkataraman,~C.; Miller,~W.~H. Semiclassical description of electronically nonadiabatic dynamics via the initial value representation. \emph{The Journal of Chemical Physics} \textbf{2007}, \emph{127}, 084114\relax
\mciteBstWouldAddEndPuncttrue
\mciteSetBstMidEndSepPunct{\mcitedefaultmidpunct}
{\mcitedefaultendpunct}{\mcitedefaultseppunct}\relax
\EndOfBibitem
\bibitem[Miller(2009)]{Miller2009}
Miller,~W.~H. Electronically Nonadiabatic Dynamics via Semiclassical Initial Value Methods. \emph{The Journal of Physical Chemistry A} \textbf{2009}, \emph{113}, 1405--1415\relax
\mciteBstWouldAddEndPuncttrue
\mciteSetBstMidEndSepPunct{\mcitedefaultmidpunct}
{\mcitedefaultendpunct}{\mcitedefaultseppunct}\relax
\EndOfBibitem
\bibitem[Miyazaki and Ananth(2023)Miyazaki, and Ananth]{Miyazaki2023}
Miyazaki,~K.; Ananth,~N. {Nonadiabatic simulations of photoisomerization and dissociation in ethylene using ab initio classical trajectories}. \emph{The Journal of Chemical Physics} \textbf{2023}, \emph{159}, 124110\relax
\mciteBstWouldAddEndPuncttrue
\mciteSetBstMidEndSepPunct{\mcitedefaultmidpunct}
{\mcitedefaultendpunct}{\mcitedefaultseppunct}\relax
\EndOfBibitem
\bibitem[NEWNS(1969)]{Newns1969}
NEWNS,~D.~M. Self-Consistent Model of Hydrogen Chemisorption. \emph{Physical Review} \textbf{1969}, \emph{178}, 1123--1135\relax
\mciteBstWouldAddEndPuncttrue
\mciteSetBstMidEndSepPunct{\mcitedefaultmidpunct}
{\mcitedefaultendpunct}{\mcitedefaultseppunct}\relax
\EndOfBibitem
\bibitem[Anderson(1961)]{Anderson1961}
Anderson,~P.~W. Localized Magnetic States in Metals. \emph{Physical Review} \textbf{1961}, \emph{124}, 41--53\relax
\mciteBstWouldAddEndPuncttrue
\mciteSetBstMidEndSepPunct{\mcitedefaultmidpunct}
{\mcitedefaultendpunct}{\mcitedefaultseppunct}\relax
\EndOfBibitem
\bibitem[Holstein(1959)]{Holstein1959}
Holstein,~T. Studies of polaron motion. \emph{Annals of Physics} \textbf{1959}, \emph{8}, 325--342\relax
\mciteBstWouldAddEndPuncttrue
\mciteSetBstMidEndSepPunct{\mcitedefaultmidpunct}
{\mcitedefaultendpunct}{\mcitedefaultseppunct}\relax
\EndOfBibitem
\bibitem[Shenvi \latin{et~al.}(2008)Shenvi, Schmidt, Edwards, and Tully]{Shenvi2008}
Shenvi,~N.; Schmidt,~J.~R.; Edwards,~S.~T.; Tully,~J.~C. {Efficient discretization of the continuum through complex contour deformation}. \emph{Physical Review A} \textbf{2008}, \emph{78}, 022502\relax
\mciteBstWouldAddEndPuncttrue
\mciteSetBstMidEndSepPunct{\mcitedefaultmidpunct}
{\mcitedefaultendpunct}{\mcitedefaultseppunct}\relax
\EndOfBibitem
\bibitem[de~Vega \latin{et~al.}(2015)de~Vega, Schollwöck, and Wolf]{deVega2015}
de~Vega,~I.; Schollwöck,~U.; Wolf,~F.~A. How to discretize a quantum bath for real-time evolution. \emph{Physical Review B} \textbf{2015}, \emph{92}, 155126\relax
\mciteBstWouldAddEndPuncttrue
\mciteSetBstMidEndSepPunct{\mcitedefaultmidpunct}
{\mcitedefaultendpunct}{\mcitedefaultseppunct}\relax
\EndOfBibitem
\bibitem[Ramchandani(1970)]{Ramchandani1970}
Ramchandani,~M.~G. {Energy band structure of gold}. \emph{Journal of Physics C: Solid State Physics} \textbf{1970}, \emph{3}, S1--S9\relax
\mciteBstWouldAddEndPuncttrue
\mciteSetBstMidEndSepPunct{\mcitedefaultmidpunct}
{\mcitedefaultendpunct}{\mcitedefaultseppunct}\relax
\EndOfBibitem
\bibitem[Kim \latin{et~al.}(2012)Kim, Kelly, Park, and Rhee]{Kim2012}
Kim,~H.~W.; Kelly,~A.; Park,~J.~W.; Rhee,~Y.~M. {All-Atom Semiclassical Dynamics Study of Quantum Coherence in Photosynthetic Fenna–Matthews–Olson Complex}. \emph{Journal of the American Chemical Society} \textbf{2012}, \emph{134}, 11640--11651\relax
\mciteBstWouldAddEndPuncttrue
\mciteSetBstMidEndSepPunct{\mcitedefaultmidpunct}
{\mcitedefaultendpunct}{\mcitedefaultseppunct}\relax
\EndOfBibitem
\bibitem[Lee \latin{et~al.}(2016)Lee, Huo, and Coker]{Lee2016}
Lee,~M.~K.; Huo,~P.; Coker,~D.~F. {Semiclassical Path Integral Dynamics: Photosynthetic Energy Transfer with Realistic Environment Interactions}. \emph{Annual Review of Physical Chemistry} \textbf{2016}, \emph{67}, 639--668\relax
\mciteBstWouldAddEndPuncttrue
\mciteSetBstMidEndSepPunct{\mcitedefaultmidpunct}
{\mcitedefaultendpunct}{\mcitedefaultseppunct}\relax
\EndOfBibitem
\bibitem[Polley and Loring(2021)Polley, and Loring]{Polley2021}
Polley,~K.; Loring,~R.~F. {Two-dimensional vibrational–electronic spectra with semiclassical mechanics}. \emph{The Journal of Chemical Physics} \textbf{2021}, \emph{154}, 194110\relax
\mciteBstWouldAddEndPuncttrue
\mciteSetBstMidEndSepPunct{\mcitedefaultmidpunct}
{\mcitedefaultendpunct}{\mcitedefaultseppunct}\relax
\EndOfBibitem
\bibitem[Polley and Loring(2022)Polley, and Loring]{Polley2022}
Polley,~K.; Loring,~R.~F. {2D electronic-vibrational spectroscopy with classical trajectories}. \emph{The Journal of Chemical Physics} \textbf{2022}, \emph{156}, 204110\relax
\mciteBstWouldAddEndPuncttrue
\mciteSetBstMidEndSepPunct{\mcitedefaultmidpunct}
{\mcitedefaultendpunct}{\mcitedefaultseppunct}\relax
\EndOfBibitem
\bibitem[Sun \latin{et~al.}(2021)Sun, Sasmal, and Vendrell]{Sun2021}
Sun,~J.; Sasmal,~S.; Vendrell,~O. A bosonic perspective on the classical mapping of fermionic quantum dynamics. \emph{The Journal of Chemical Physics} \textbf{2021}, \emph{155}, 134110\relax
\mciteBstWouldAddEndPuncttrue
\mciteSetBstMidEndSepPunct{\mcitedefaultmidpunct}
{\mcitedefaultendpunct}{\mcitedefaultseppunct}\relax
\EndOfBibitem
\bibitem[Montoya-Castillo and Markland(2023)Montoya-Castillo, and Markland]{Montoya2023}
Montoya-Castillo,~A.; Markland,~T.~E. A derivation of the conditions under which bosonic operators exactly capture fermionic structure and dynamics. \emph{The Journal of Chemical Physics} \textbf{2023}, \emph{158}, 094112\relax
\mciteBstWouldAddEndPuncttrue
\mciteSetBstMidEndSepPunct{\mcitedefaultmidpunct}
{\mcitedefaultendpunct}{\mcitedefaultseppunct}\relax
\EndOfBibitem
\bibitem[Jung \latin{et~al.}(2023)Jung, Kelly, and Markland]{Jung2023}
Jung,~K.~A.; Kelly,~J.; Markland,~T.~E. Electron transfer at electrode interfaces via a straightforward quasiclassical fermionic mapping approach. \emph{The Journal of Chemical Physics} \textbf{2023}, \emph{159}\relax
\mciteBstWouldAddEndPuncttrue
\mciteSetBstMidEndSepPunct{\mcitedefaultmidpunct}
{\mcitedefaultendpunct}{\mcitedefaultseppunct}\relax
\EndOfBibitem
\bibitem[Xu \latin{et~al.}(2019)Xu, Liu, Song, and Shi]{Xu2019}
Xu,~M.; Liu,~Y.; Song,~K.; Shi,~Q. A non-perturbative approach to simulate heterogeneous electron transfer dynamics: Effective mode treatment of the continuum electronic states. \emph{The Journal of Chemical Physics} \textbf{2019}, \emph{150}\relax
\mciteBstWouldAddEndPuncttrue
\mciteSetBstMidEndSepPunct{\mcitedefaultmidpunct}
{\mcitedefaultendpunct}{\mcitedefaultseppunct}\relax
\EndOfBibitem
\bibitem[Dan and Shi(2023)Dan, and Shi]{Dan2023}
Dan,~X.; Shi,~Q. Theoretical study of nonadiabatic hydrogen atom scattering dynamics on metal surfaces using the hierarchical equations of motion method. \emph{The Journal of Chemical Physics} \textbf{2023}, \emph{159}\relax
\mciteBstWouldAddEndPuncttrue
\mciteSetBstMidEndSepPunct{\mcitedefaultmidpunct}
{\mcitedefaultendpunct}{\mcitedefaultseppunct}\relax
\EndOfBibitem
\end{mcitethebibliography}


\providecommand{\latin}[1]{#1}
\makeatletter
\providecommand{\doi}
  {\begingroup\let\do\@makeother\dospecials
  \catcode`\{=1 \catcode`\}=2 \doi@aux}
\providecommand{\doi@aux}[1]{\endgroup\texttt{#1}}
\makeatother
\providecommand*\mcitethebibliography{\thebibliography}
\csname @ifundefined\endcsname{endmcitethebibliography}  {\let\endmcitethebibliography\endthebibliography}{}
\begin{mcitethebibliography}{19}
\providecommand*\natexlab[1]{#1}
\providecommand*\mciteSetBstSublistMode[1]{}
\providecommand*\mciteSetBstMaxWidthForm[2]{}
\providecommand*\mciteBstWouldAddEndPuncttrue
  {\def\EndOfBibitem{\unskip.}}
\providecommand*\mciteBstWouldAddEndPunctfalse
  {\let\EndOfBibitem\relax}
\providecommand*\mciteSetBstMidEndSepPunct[3]{}
\providecommand*\mciteSetBstSublistLabelBeginEnd[3]{}
\providecommand*\EndOfBibitem{}
\mciteSetBstSublistMode{f}
\mciteSetBstMaxWidthForm{subitem}{(\alph{mcitesubitemcount})}
\mciteSetBstSublistLabelBeginEnd
  {\mcitemaxwidthsubitemform\space}
  {\relax}
  {\relax}

\bibitem[Gardner \latin{et~al.}(2023)Gardner, Habershon, and Maurer]{Gardner2023}
Gardner,~J.; Habershon,~S.; Maurer,~R.~J. {Assessing Mixed Quantum-Classical Molecular Dynamics Methods for Nonadiabatic Dynamics of Molecules on Metal Surfaces}. \emph{The Journal of Physical Chemistry C} \textbf{2023}, \relax
\mciteBstWouldAddEndPunctfalse
\mciteSetBstMidEndSepPunct{\mcitedefaultmidpunct}
{}{\mcitedefaultseppunct}\relax
\EndOfBibitem
\bibitem[Meng and Jiang(2022)Meng, and Jiang]{Meng2022}
Meng,~G.; Jiang,~B. {A pragmatic protocol for determining charge transfer states of molecules at metal surfaces by constrained density functional theory}. \emph{The Journal of Chemical Physics} \textbf{2022}, \emph{157}, 214103\relax
\mciteBstWouldAddEndPuncttrue
\mciteSetBstMidEndSepPunct{\mcitedefaultmidpunct}
{\mcitedefaultendpunct}{\mcitedefaultseppunct}\relax
\EndOfBibitem
\bibitem[Meyer and Miller(1979)Meyer, and Miller]{Meyer1979}
Meyer,~H.-D.; Miller,~W.~H. A classical analog for electronic degrees of freedom in nonadiabatic collision processes. \emph{The Journal of Chemical Physics} \textbf{1979}, \emph{70}, 3214--3223\relax
\mciteBstWouldAddEndPuncttrue
\mciteSetBstMidEndSepPunct{\mcitedefaultmidpunct}
{\mcitedefaultendpunct}{\mcitedefaultseppunct}\relax
\EndOfBibitem
\bibitem[Stock and Thoss(1997)Stock, and Thoss]{Stock1997}
Stock,~G.; Thoss,~M. Semiclassical Description of Nonadiabatic Quantum Dynamics. \emph{Physical Review Letters} \textbf{1997}, \emph{78}, 578\relax
\mciteBstWouldAddEndPuncttrue
\mciteSetBstMidEndSepPunct{\mcitedefaultmidpunct}
{\mcitedefaultendpunct}{\mcitedefaultseppunct}\relax
\EndOfBibitem
\bibitem[Ramchandani(1970)]{Ramchandani1970}
Ramchandani,~M.~G. {Energy band structure of gold}. \emph{Journal of Physics C: Solid State Physics} \textbf{1970}, \emph{3}, S1--S9\relax
\mciteBstWouldAddEndPuncttrue
\mciteSetBstMidEndSepPunct{\mcitedefaultmidpunct}
{\mcitedefaultendpunct}{\mcitedefaultseppunct}\relax
\EndOfBibitem
\bibitem[Shenvi \latin{et~al.}(2009)Shenvi, Roy, and Tully]{Shenvi2009}
Shenvi,~N.; Roy,~S.; Tully,~J.~C. {Nonadiabatic dynamics at metal surfaces: Independent-electron surface hopping}. \emph{The Journal of Chemical Physics} \textbf{2009}, \emph{130}, 174107\relax
\mciteBstWouldAddEndPuncttrue
\mciteSetBstMidEndSepPunct{\mcitedefaultmidpunct}
{\mcitedefaultendpunct}{\mcitedefaultseppunct}\relax
\EndOfBibitem
\bibitem[Gardner \latin{et~al.}(2023)Gardner, Corken, Janke, Al, Habershon, and Maurer]{Gardner2023b}
Gardner,~J.; Corken,~D.; Janke,~S.~M.; Al; Habershon,~S.; Maurer,~R.~J. {Efficient implementation and performance analysis of the independent electron surface hopping method for dynamics at metal surfaces}. \emph{The Journal of Chemical Physics} \textbf{2023}, \emph{158}, 64101\relax
\mciteBstWouldAddEndPuncttrue
\mciteSetBstMidEndSepPunct{\mcitedefaultmidpunct}
{\mcitedefaultendpunct}{\mcitedefaultseppunct}\relax
\EndOfBibitem
\bibitem[Frank \latin{et~al.}(2000)Frank, Rivera, and Wolf]{Frank2000}
Frank,~A.; Rivera,~A.~L.; Wolf,~K.~B. Wigner function of Morse potential eigenstates. \emph{Physical Review A} \textbf{2000}, \emph{61}, 054102\relax
\mciteBstWouldAddEndPuncttrue
\mciteSetBstMidEndSepPunct{\mcitedefaultmidpunct}
{\mcitedefaultendpunct}{\mcitedefaultseppunct}\relax
\EndOfBibitem
\bibitem[Yamamoto \latin{et~al.}(2002)Yamamoto, Wang, and Miller]{Yamamoto2002c}
Yamamoto,~T.; Wang,~H.; Miller,~W.~H. Combining semiclassical time evolution and quantum Boltzmann operator to evaluate reactive flux correlation function for thermal rate constants of complex systems. \emph{The Journal of Chemical Physics} \textbf{2002}, \emph{116}, 7335--7349\relax
\mciteBstWouldAddEndPuncttrue
\mciteSetBstMidEndSepPunct{\mcitedefaultmidpunct}
{\mcitedefaultendpunct}{\mcitedefaultseppunct}\relax
\EndOfBibitem
\bibitem[Frenkel and Smit(2002)Frenkel, and Smit]{FRENKEL200223}
Frenkel,~D.; Smit,~B. In \emph{Understanding Molecular Simulation (Second Edition)}, second edition ed.; Frenkel,~D., Smit,~B., Eds.; Academic Press: San Diego, 2002; pp 23--61\relax
\mciteBstWouldAddEndPuncttrue
\mciteSetBstMidEndSepPunct{\mcitedefaultmidpunct}
{\mcitedefaultendpunct}{\mcitedefaultseppunct}\relax
\EndOfBibitem
\bibitem[Swenson \latin{et~al.}(2011)Swenson, Levy, Cohen, Rabani, and Miller]{Swenson2011}
Swenson,~D. W.~H.; Levy,~T.; Cohen,~G.; Rabani,~E.; Miller,~W.~H. Application of a semiclassical model for the second-quantized many-electron Hamiltonian to nonequilibrium quantum transport: The resonant level model. \emph{The Journal of Chemical Physics} \textbf{2011}, \emph{134}\relax
\mciteBstWouldAddEndPuncttrue
\mciteSetBstMidEndSepPunct{\mcitedefaultmidpunct}
{\mcitedefaultendpunct}{\mcitedefaultseppunct}\relax
\EndOfBibitem
\bibitem[Li \latin{et~al.}(2013)Li, Levy, Swenson, Rabani, and Miller]{Li2013}
Li,~B.; Levy,~T.~J.; Swenson,~D. W.~H.; Rabani,~E.; Miller,~W.~H. A Cartesian quasi-classical model to nonequilibrium quantum transport: The Anderson impurity model. \emph{The Journal of Chemical Physics} \textbf{2013}, \emph{138}, 104110\relax
\mciteBstWouldAddEndPuncttrue
\mciteSetBstMidEndSepPunct{\mcitedefaultmidpunct}
{\mcitedefaultendpunct}{\mcitedefaultseppunct}\relax
\EndOfBibitem
\bibitem[Levy \latin{et~al.}(2019)Levy, Dou, Rabani, and Limmer]{Levy2019}
Levy,~A.; Dou,~W.; Rabani,~E.; Limmer,~D.~T. A complete quasiclassical map for the dynamics of interacting fermions. \emph{The Journal of Chemical Physics} \textbf{2019}, \emph{150}, 234112\relax
\mciteBstWouldAddEndPuncttrue
\mciteSetBstMidEndSepPunct{\mcitedefaultmidpunct}
{\mcitedefaultendpunct}{\mcitedefaultseppunct}\relax
\EndOfBibitem
\bibitem[Jung \latin{et~al.}(2023)Jung, Kelly, and Markland]{Jung2023}
Jung,~K.~A.; Kelly,~J.; Markland,~T.~E. Electron transfer at electrode interfaces via a straightforward quasiclassical fermionic mapping approach. \emph{The Journal of Chemical Physics} \textbf{2023}, \emph{159}\relax
\mciteBstWouldAddEndPuncttrue
\mciteSetBstMidEndSepPunct{\mcitedefaultmidpunct}
{\mcitedefaultendpunct}{\mcitedefaultseppunct}\relax
\EndOfBibitem
\bibitem[Bonella and Coker(2003)Bonella, and Coker]{Bonella2003}
Bonella,~S.; Coker,~D.~F. Semiclassical implementation of the mapping Hamiltonian approach for nonadiabatic dynamics using focused initial distribution sampling. \emph{The Journal of Chemical Physics} \textbf{2003}, \emph{118}, 4370--4385\relax
\mciteBstWouldAddEndPuncttrue
\mciteSetBstMidEndSepPunct{\mcitedefaultmidpunct}
{\mcitedefaultendpunct}{\mcitedefaultseppunct}\relax
\EndOfBibitem
\bibitem[Runeson and Richardson(2019)Runeson, and Richardson]{Runeson2019}
Runeson,~J.~E.; Richardson,~J.~O. Spin-mapping approach for nonadiabatic molecular dynamics. \emph{The Journal of Chemical Physics} \textbf{2019}, \emph{151}, 044119\relax
\mciteBstWouldAddEndPuncttrue
\mciteSetBstMidEndSepPunct{\mcitedefaultmidpunct}
{\mcitedefaultendpunct}{\mcitedefaultseppunct}\relax
\EndOfBibitem
\bibitem[Kelly \latin{et~al.}(2012)Kelly, van Zon, Schofield, and Kapral]{Kelly2012}
Kelly,~A.; van Zon,~R.; Schofield,~J.; Kapral,~R. Mapping quantum-classical Liouville equation: Projectors and trajectories. \emph{The Journal of Chemical Physics} \textbf{2012}, \emph{136}, 084101\relax
\mciteBstWouldAddEndPuncttrue
\mciteSetBstMidEndSepPunct{\mcitedefaultmidpunct}
{\mcitedefaultendpunct}{\mcitedefaultseppunct}\relax
\EndOfBibitem
\bibitem[Bonella and Coker(2001)Bonella, and Coker]{Bonella2001}
Bonella,~S.; Coker,~D.~F. A semiclassical limit for the mapping Hamiltonian approach to electronically nonadiabatic dynamics. \emph{The Journal of Chemical Physics} \textbf{2001}, \emph{114}, 7778--7789\relax
\mciteBstWouldAddEndPuncttrue
\mciteSetBstMidEndSepPunct{\mcitedefaultmidpunct}
{\mcitedefaultendpunct}{\mcitedefaultseppunct}\relax
\EndOfBibitem
\end{mcitethebibliography}

\end{document}


\maketitle

\newpage

\section{Refitting the Model Potential}\label{sec:refit}
As mentioned in the main text, the model presented by \citeauthor{Gardner2023} uses simplified functional forms to fit diabats for the NO and NO$^-$ states, Eqs.(3) and (4), and the adiabatic ground state of the NAH Hamiltonian, Eq.(1)  to constrained DFT data.\cite{Meng2022} For the metal states, they use a wide band approximation with a band width $\Delta E = 100$ eV to obtain the ground state of the NAH Hamiltonian. In the LSC method, the electronic states of the metal are mapped onto fictitious harmonic oscillators as prescribed in the MMST mapping\cite{Meyer1979,Stock1997}. The frequency of these oscillators depends on the energy of the electronic state. For a wide band of band width 100 ev, many electronic states on the edge of the band will correspond to a very high energy and will require a very small time step to numerically integrate Hamilton's equations of motion, making the use of many trajectories prohibitively expensive. Moreover, the LSC method doesn't require the use of a wide band, unlike other methods explored by \citeauthor{Gardner2023}, and in the interest of better capturing the physics, we reparameterize their model using a more realistic band width for gold of $\Delta E = 7$ eV.\cite{Ramchandani1970,Shenvi2009}

\begin{figure}
     \centering
     \begin{subfigure}[b]{0.32\textwidth}
         \centering
         \includegraphics[width=\textwidth]{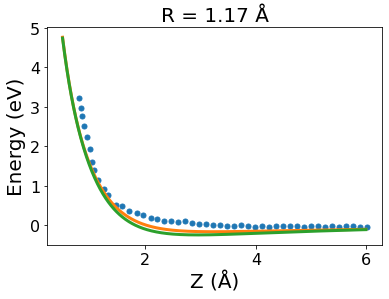}
     \end{subfigure}
     \hfill
     \begin{subfigure}[b]{0.32\textwidth}
         \centering
         \includegraphics[width=\textwidth]{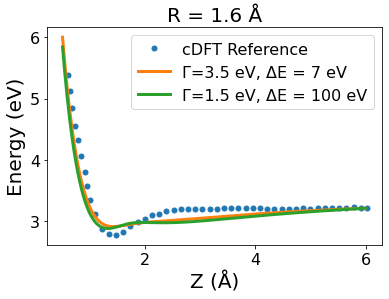}
         \end{subfigure}
     \hfill
     \begin{subfigure}[b]{0.32\textwidth}
         \centering
         \includegraphics[width=\textwidth]{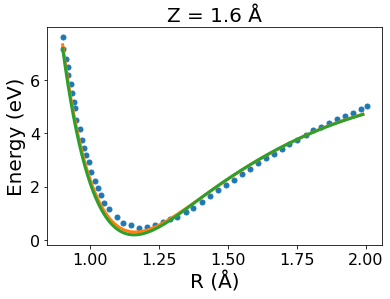}
        \end{subfigure}
        \caption{Slices of the ground state adiabatic potential energy surface for the NAH Hamiltonian comparing the fit of \citeauthor{Gardner2023} $(\Gamma = 1.5 \text{eV}, \Delta E = 100 \text{eV})$ and our refit $(\Gamma = 3.5 \text{eV}, \Delta E = 7 \text{eV})$ with cDFT reference data\cite{Meng2022} (obtained via private communication).}
        \label{fig:refit}
\end{figure}
\begin{figure}
     \centering
     \begin{subfigure}[b]{0.32\textwidth}
         \centering
         \includegraphics[width=\textwidth]{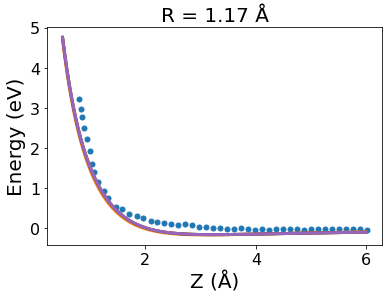}
     \end{subfigure}
     \hfill
     \begin{subfigure}[b]{0.32\textwidth}
         \centering
         \includegraphics[width=\textwidth]{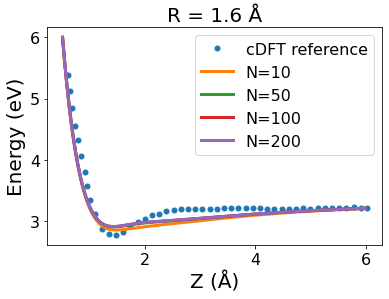}
         \end{subfigure}
     \hfill
     \begin{subfigure}[b]{0.32\textwidth}
         \centering
         \includegraphics[width=\textwidth]{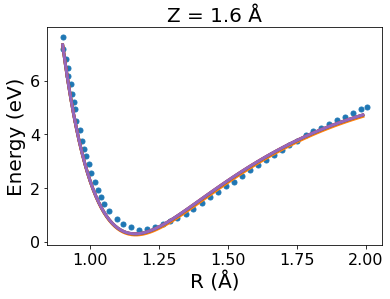}
        \end{subfigure}
        \caption{Slices of the adiabatic ground state energy for the refit potential $(\Gamma = 3.5 \text{eV}, \Delta E = 7 \text{eV})$ with varying number of metal bath states (N), along with the reference cDFT data\cite{Meng2022} (obtained via private communication). Plots for N = 50, 100 and 200 are overlapping for all three slices.}
        \label{fig:refit_Nbconv}
\end{figure}

\begin{figure}
     \centering
     \begin{subfigure}[b]{0.49\textwidth}
         \centering
         \includegraphics[width=\textwidth]{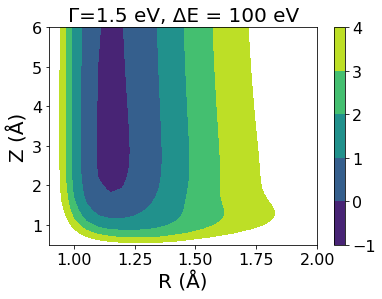}
     \end{subfigure}
     \hfill
     \begin{subfigure}[b]{0.49\textwidth}
         \centering
         \includegraphics[width=\textwidth]{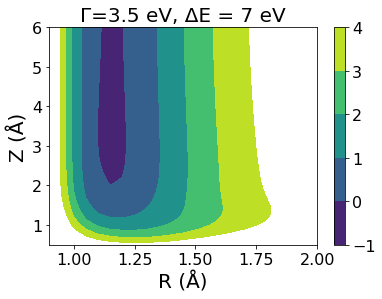}
         \end{subfigure}
     \hfill
     \caption{Contour plots of the adiabatic ground state energy for the model of \citeauthor{Gardner2023} $(\Gamma = 1.5 \text{eV}, \Delta E = 100 \text{eV})$ and our refit potential $(\Gamma = 3.5 \text{eV}, \Delta E = 7 \text{eV})$.}
    \label{fig:refit_contours}
\end{figure}

We choose to use the same functional forms and parameters for the two diabats $U_{0}(\mathbf{X})$ and  $U_{1}(\mathbf{X})$. In order to fit the ground state adiabatic energy $E_{0}(\mathbf{X})$ using our new band width, we calculate the ground state energy as,\cite{Gardner2023b}
$E_0 = U_0 + \sum_{k=1}^{N} f(\lambda_k)\lambda_k$, where $\{\lambda_k\}$ are the eigenvalues of the NAH Hamiltonian and $f(\lambda_k)$ is the fermi function that determines the occupation of the energy level $\lambda_k$. To ensure that the reference electronic structure data and the calculated $E_0$ have the same zero of energy, we set $E_0(\mathbf{X}_{ref}) = E_0^{cDFT}(\mathbf{X}_{ref})$ at a reference geometry $R_{ref} = 1.6\AA, \, Z_{ref} = 6.02\AA$. We get a refit that closely resembles the fit of \citeauthor{Gardner2023}, as shown in Fig.~\ref{fig:refit}. Fig.~\ref{fig:refit_Nbconv} shows that the fitting procedure has converged with respect to the number of metal states M. Fig.~\ref{fig:refit_contours} shows a contour plot of the adiabatic ground state  energy as a function of the nuclear dofs R and Z.

\section{Simulation Details}
\subsection{Initial Conditions} \label{subsec:initcon}
As mentioned in the main text, the initial density is chosen to be $\hat{\rho} = \ketbra{P_{Zi},Z_i}{P_{Zi},Z_i}\otimes\ketbra{\nu_i}{\nu_i}\otimes \hat{\rho}_{\text{mol}} \otimes \hat{\rho}_{\text{metal}}$. The translational dof is projected onto the coherent state $\braket{Z}{P_{Zi},Z_i} = \left(\frac{\gamma_Z}{\pi}\right)^{1/4}e^{-\frac{\gamma_Z}{2}\left(Z-Z_i\right)^2 + iP_{Zi}(Z-Z_i)/\hbar}$ centered at $Z_i = 5\AA$, $P_{Zi}= -\sqrt{2 m_Z E_i}$, where $m_Z$ is the mass of the NO molecule and $E_i$ is the incident translational energy. The negative sign in the momentum ensures that the molecule is directed towards the surface, and the width of the coherent state is chosen as $\gamma_Z = 4.544 \AA^{-2}$. Since the NO molecule is initialized far from the surface ($Z_i = 5\AA$), the potential in the R dof is a Morse oscillator corresponding to the neutral NO molecule, with the mass taken as the reduced mass of the NO molecule. We initialize the R dof by projecting it onto the $\nu_i^{th}$ eigenstate of the Morse oscillator $\ket{\nu_i}$. Upon initialization, the NO molecule is in its neutral state, and thus the NO$^-$ molecular state is unoccupied, $\hat{\rho}_{\text{mol}} = \hat{d}\hat{d}^{\dag}$. The metal is in thermal and chemical equilibrium at inverse temperature $\beta$ and chemical potential $\mu$, $\hat{\rho}_{\text{metal}} = \prod_{k=1}^{N}\frac{e^{-\beta\left(\epsilon_k -\mu\right)\hat{c}_k^{\dag}\hat{c}_k}}{\text{Tr}_k\left[e^{-\beta\left(\epsilon_k -\mu\right)\hat{c}_k^{\dag}\hat{c}_k}\right]}$. Throughout this study, we set $\mu = 0$.

To calculate expectation values of a general operator $\hat{B}$ using Eq.(7), we need to perform the $2(F+D)$ dimensional integral using Monte Carlo(MC) sampling. The translational coordinates $(Z,P_Z)$ are sampled as Gaussian random variables  centered at $(Z=Z_i,P_Z = P_{Zi})$ from $\left[\ketbra{P_{Zi},Z_i}{P_{Zi},Z_i}\right]_W(Z,P_Z)$. Sampling the vibrational coordinates $(R,P_R)$ is a little less straightforward. Although the Wigner transform $\left[\ketbra{\nu_i}{\nu_i}\right]_W(R,P_R)$ is known analytically, it is challenging to use as a sampling function.\cite{Frank2000} So we choose to calculate this Wigner transform on a sinc-DVR grid as presented in Ref.\citenum{Yamamoto2002c} and sample on its absolute magnitude using a Metropolis MC procedure.\cite{FRENKEL200223} The MC step size is chosen so that 40-60\% of proposed MC steps are accepted. Periodic boundary conditions are used while performing MC steps to keep $(R,P_R)$ on the DVR grid. The DVR grid spans $R \in [0.9 \AA, 2.4 \AA]$, with 75 grid points. The range and density of the grid are chosen such that various different values of $(R,P_R)$ seen over the course of real-time propagation are within the range of the DVR grid. For the mapped electronic dofs, $(x_k,p_k)$, $k \in [0,N]$, we do not sample on the Wigner transform of the density $\rho_{Mol}\otimes\rho_{Metal}$, but instead take a more approximate approach.\cite{Swenson2011,Li2013,Levy2019,Jung2023} The metal states are at thermal equilibrium, implying that the population of each state is given by the fermi distribution, $\langle n_k \rangle = f(\epsilon_k) = (1 + e^{\beta(\mu-\epsilon_k)})^{-1}$ for $k \in [1,N]$. We can ensure that each state (for instance the $k^{th}$) has the correct thermally averaged occupation by choosing to keep it occupied $(n_k = 1)$ with a probability $f(\epsilon_k)$, and unoccupied $(n_k = 0)$ otherwise. This can be implemented by generating a random number $\xi$ and choosing the occupation number for the $k^{th}$ state as follows,
\begin{equation}
    n_k = 
    \begin{cases}
        1 & \text{if } \xi \le (1 + e^{\beta(\mu-\epsilon_k)})^{-1}\\
        0 & \text{if } \xi > (1 + e^{\beta(\mu-\epsilon_k)})^{-1}.
    \end{cases}
\end{equation}
The molecular state is initially unoccupied, $\langle n_0 \rangle = 0$, and we keep it unoccupied for each sampled point, $n_0 = 0$. Once the occupation number for each mapped state has been determined, we use focused sampling to sample $\left( x_k, p_k \right)$ on a circle in phase space such that, $\frac{1}{2}(x_k^2 + p_k^2 - \gamma) = n_k$.\cite{Bonella2003} In the standard MMST mapping, the ZPE parameter, $\gamma = 1$. Unfortunately, approximate dynamic methods like LSC do not conserve ZPE in mapped oscillators, and can result in negative electronic populations. We observe this in our simulations when we use $\gamma = 1$. Instead, motivated by the spin mappping approach, we set $\gamma =0$.~\cite{Runeson2019} With the use of a symmetrized Hamiltonian and focused sampling, the operators corresponding to electronic populations in MMST mapping and the more recently proposed spin-mapping only differ in the choice of the ZPE parameter $\gamma$. In the spin mapping scheme, $\gamma \to 0$ as the number of mapped states $N+1 \to \infty$, which is the case in the NAH Hamiltonian. 

\subsection{Dynamics} \label{subsec:dyn}
Using the MMST mapping, the NAH Hamiltonian, Eq.(1) can be mapped onto the classical hamiltonian, 
\begin{align}
   {H}_{map}({\mathbf{X}},{\mathbf{P}},{\mathbf{x}},{\mathbf{p}}) &= \frac{1}{2}{\mathbf{P}}.\mathbf{M}^{-1}.{\mathbf{P}} + U_{0}({\mathbf{X}}) + \frac{1}{2}h({\mathbf{X}})({x}_0^2+{p}_0^2-\gamma) + \Sigma_{k=1}^{N} \frac{\epsilon_{k}}{2}({x}_k^2+{p}_k^2-\gamma)  \notag \\
    &+ \Sigma_{k=1}^{N} V_{k}({\mathbf{X}})({x}_0{x}_k + {p}_0{p}_k), \label{NAH-map}
\end{align}
We chose to symmetrize the Hamiltonian so that the dynamics is independent of $\gamma$.\cite{Kelly2012} The mapped classical Hamiltonian, Eq.~\eqref{NAH-map} can be rewritten as, 
\begin{align}
     {H}_{map}({\mathbf{X}},{\mathbf{P}},{\mathbf{x}},{\mathbf{p}}) &= \frac{1}{2}{\mathbf{P}}.\mathbf{M}^{-1}.{\mathbf{P}} + U_{0}({\mathbf{X}}) + \frac{1}{2}\left(\mathbf{x}.\mathbf{V}({\mathbf{X}}).\mathbf{x} + \mathbf{p}.\mathbf{V}({\mathbf{X}}).\mathbf{p} -\gamma\text{Tr}\left[\mathbf{V({\mathbf{X}})}\right]\right), \label{h-map-cl}
\end{align}
where $\mathbf{V}(\mathbf{X})$ is defined as
\begin{align}
    \mathbf{V}({\mathbf{X}}) = \left(\begin{array}{ccccc}
    h(\mathbf{X}) & V_1(\mathbf{X}) & V_2(\mathbf{X}) & \hdots & V_M(\mathbf{X}) \\
    V_1(\mathbf{X}) & \epsilon_1 &  0 & \hdots & \hdots\\
    V_2(\mathbf{X}) & 0 & \epsilon_2 & \ddots & \vdots \\ 
    \vdots & \vdots & \ddots & \ddots & \vdots \\
    V_M(\mathbf{X}) & \vdots & \hdots & \hdots &\epsilon_N  
    \end{array}\right)_{(N+1) \times (N+1)},
\end{align}
and the vector $\mathbf{x}$ is defined as $\mathbf{x} = \left(x_0, x_1,...,x_M\right)$ and similarly for $\mathbf{p}$. We note that the total number of excitations in the mapped oscillators corresponds to the total number of electrons in the metal states $(N_e)$. To symmetrize the Hamiltonian, we add and subtract Tr$[\mathbf{V}]N_e/(N+1)\mathbb{1}$, and use, $ N_e = \sum_{k=0}^{N}\frac{1}{2}\left(x_k^2 + p_k^2 - \gamma\right)$, which on some simplification yeilds,
\begin{align}
    {H}_{sym}({\mathbf{X}},{\mathbf{P}},{\mathbf{x}},{\mathbf{p}}) &= \frac{1}{2}{\mathbf{P}}.\mathbf{M}^{-1}.{\mathbf{P}} + \tilde{U}({\mathbf{X}}) + \frac{1}{2}\left[\mathbf{x}.\mathbf{\tilde{V}}({\mathbf{X}}).\mathbf{x} + \mathbf{p}.\mathbf{\tilde{V}}({\mathbf{X}}).\mathbf{p} \right], 
\end{align}
where $\tilde{U}({\mathbf{X}}) = U_0({\mathbf{X}}) + \frac{N_e}{N+1}\text{Tr}[V({\mathbf{X}})]$ and $\mathbf{\tilde{V}}({\mathbf{X}}) =\mathbf{V}({\mathbf{X}}) - \frac{1}{N+1}\text{Tr}[V({\mathbf{X}})]\mathbb{1}$. Classical trajectories are run under the classical Hamiltonian $H_{sym}$. A time step of $1.4 - 1.6\times10^{-2}$ fs is used for the various different incident energies. To check for energy conservation, trajectories not obeying the condition, $\frac{E(t)-E(0)}{E(0)} \le 10^{-3}$ are discarded. Trajectories not obeying $ 0.5 \text{ a.u.} \le R \le  10  \text{ a.u.}$ are discarded to ensure that they are not subjected to an inverted potential.\cite{Bonella2001} In total, $<0.1\%$ of the trajectories are discarded. Trajectories are propagated for a sufficiently long time to ensure that the final vibrational state probability reaches a plateau value. Error bars for vibrational transition probabilities are obtained by averaging over any remaining small amplitude oscillations at longer times. 

Lastly, operators corresponding to all the observables calculated in this study are presented in Table \ref{tbl:oper}, along with their Wigner transforms.

\begin{table}
    \caption{Operators corresponding to various observables calculated in this study and their respective wigner transforms.}
  \label{tbl:oper}
 
  \begin{tabular}{|c|c|c|}
  \hline
    Observable   &  Operator    &  Wigner Transform  \\
    \hline \hline 
    & &  \\
    NO bond length & $\hat{R}$ & R \\ \hline
    Distance of NO center of mass & & \\
     from the surface      & $\hat{Z}$    & $Z$   \\ \hline
     & &  \\
    Population of NO$^-$ state      &  $\hat{d}^{\dag}\hat{d} \to \frac{1}{2}\left(\hat{x}_0^2 + \hat{p}_0^2 - \gamma \right)$  & $\frac{1}{2}\left(x_0^2 + {p}_0^2 - \gamma \right)$   \\ \hline
    & &  \\
    Population of $k^{th}$ metal state       & $\hat{c}_{k}^{\dag}\hat{c}_k \to \frac{1}{2}\left(\hat{x}_k^2 + \hat{p}_k^2 - \gamma \right)$    & $\frac{1}{2}\left(x_k^2 + {p}_k^2 - \gamma \right)$ \\ \hline
    & & $\left[\ketbra{\nu_i}{\nu_i}\right]_W(R,P_R)$  \\ 
    Population of $\nu_i^{th}$ vibrational state & $\ketbra{\nu_i}{\nu_i}$ &  calculated on a DVR grid\\
    \hline \hline
  \end{tabular}
  
\end{table}
\bibliography{zbibfile_si}